\documentclass[11pt]{article}
\usepackage{amsmath,amssymb,color,comment,graphics,epsfig,cite}
\usepackage{amsfonts}
\usepackage{hyperref}
\usepackage{multirow}
\usepackage{booktabs}
\usepackage{amsmath}
\usepackage{subcaption}
\newcommand{\hoch}[1]{$\, ^{#1}$}
\newcommand{\be}{\begin{equation}}
\newcommand{\ee}{\end{equation}}
\newcommand{\bea}{\setlength\arraycolsep{2pt} \begin{eqnarray}}
\newcommand{\eea}{\end{eqnarray}}
\newcommand{\nn}{\nonumber}
\def\a{\alpha}

\def\m{\mu}
\def\n{\nu}

\def\s{\sigma}

\def\O{\Omega}

\providecommand{\keywords}[1]
{
  \small	
  \textbf{\textit{Keywords---}} #1
}

\begin{document}

\begin{center}
{\Large {\bf Full spectrum of Love numbers of Reissner-Nordstr\"om  black hole in $D$-dimensions }}

\vspace{20pt}

Minghao Xia\hoch{1}, Liang Ma\hoch{1}, Yi Pang\hoch{1,2} and  Hong. L\"{u}\hoch{1,2,3}

\vspace{10pt}

{\it \hoch{1}Center for Joint Quantum Studies and Department of Physics,\\
School of Science, Tianjin University, Tianjin 300350, China }

\bigskip

{\it \hoch{2}Peng Huanwu Center for Fundamental Theory, Hefei, Anhui 230026, China}

\bigskip

{\it \hoch{3}Joint School of National University of Singapore and Tianjin University,\\
International Campus of Tianjin University, Binhai New City, Fuzhou 350207, China}

\vspace{40pt}

\underline{ABSTRACT}
\end{center}

We present a comprehensive analysis of the full spectrum of tidal Love numbers for Reissner-Nordstr\"om (RN) black holes in general spacetime dimensions. By perturbing the Einstein-Maxwell theory around the $D$-dimensional RN background, we derive an effective two dimensional quadratic action encompassing tensor, vector, and scalar-type perturbation sectors. Through diagonalization, we obtain master equations governing each sector and extract the corresponding Love numbers from the asymptotic behavior of the solutions. Our results confirm that all Love numbers vanish for four-dimensional RN black holes. In higher dimensions, the tensor and vector Love numbers reproduce previously known results. For the previously unknown scalar-type Love numbers, we show also they vanish for integer valued effective multipolar indices and display logarithmic running behavior when the corresponding indices are half integers.


\vfill {\footnotesize  pangyi1@tju.edu.cn (corresponding author)
}

\thispagestyle{empty}
\pagebreak
\tableofcontents
\addtocontents{toc}{\protect\setcounter{tocdepth}{2}}

\section{Introduction}
Tidal Love numbers describe the deformability of a black hole under the influence of external force. It has been demonstrated that conservative linear tidal Love numbers vanish for several families of asymptotically flat black holes in four dimensions, including Schwarzschild \cite{Binnington:2009bb}, Kerr \cite{LeTiec:2020bos, Chia:2020yla, Charalambous:2021mea} and  Reissner-Nordstr\"om (RN) \cite{Cardoso:2017cfl, Pereniguez:2021xcj, Rai:2024lho} black holes. It was also found that for Kerr black holes, the conservative tidal Love numbers remain vanishing at nonlinear level
\cite{DeLuca:2023mio, Riva:2023rcm, Sharma:2024hlz, Gounis:2024hcm}. These results provide strong support to the no-hair theorem, showing that black holes in four-dimensional   general relativity are remarkably rigid objects. On the other hand, Love numbers are generically nonvanishing for dynamical black holes \cite{Bhatt:2023zsy,Bhatt:2024yyz,Bhatt:2024rpx}, charged black holes coupled to charged scalar or fermion \cite{Ma:2024few,Pereniguez:2025jxq ,Chakraborty:2025zyb, Pang:2025myy }
and
black holes in higher dimensions
\cite{Rodriguez:2023xjd,Cardoso:2019vof, Charalambous:2023jgq, Glazer:2024eyi}, in modified gravitational theories \cite{Cardoso:2018ptl,DeLuca:2022tkm,Barura:2024uog, Katagiri:2024fpn,Barbosa:2025uau ,Motaharfar:2025ihv,Motaharfar:2025typ}, in the presence of a cosmological constant \cite{Nair:2024mya, Franzin:2024cah}, or in astrophysical environments \cite{Cardoso:2019upw,DeLuca:2021ite,DeLuca:2022xlz,Katagiri:2023yzm,Cannizzaro:2024fpz}. Therefore, Love numbers are useful theoretical diagnostics. In fundamental theories, such as string theory or supergravit theory, Love numbers might be used to distinguish black holes from other compact objects, such as the black rings \cite{Emparan:2001wn} or fuzzball solutions \cite{Mathur:2005zp}. Potential detectability by future gravitational wave experiments make Love numbers even
more intriguing physical observables.

In this work, we shall revisit the computation  of Love numbers of RN  black holes in diverse dimensions for several reasons. In the previous work \cite{Pereniguez:2021xcj}, only tensor and vector Love numbers were investigated. However, to gain complete knowledge about how RN black holes respond to external sources, it is necessary to compute also the scalar type Love numbers. On the other hand, within the framework of the realistic Einstein-Maxwell theory, charged black holes represent the simplest generalizations of the vacuum Schwarzschild solution. It provides a natural testbed to study graviton-photon mixing \cite{mix1,mix2}, which may lead to conversion of gravitational perturbations into their electromagnetic counterparts and vice versa.  Although astrophysical black holes are usually assumed to be neutral, there are well known astrophysical processes by which black holes can acquire charge \cite{Zajacek:2018ycb, Levin:2018mzg}, albeit typically small.

We carry out the analysis by expanding the $D$-dimensional fields in terms of harmonic functions on $S^{D-2}$, plugging them into the Einstein-Maxwell action,
linearized about the background RN black hole to yield an effective 2-dimensional quadratic
action for perturbative fields in $1+1$ dimensions. Upon diagonalizing the perturbations, one obtains the complete set of master equations generalizing previous results for Schwarzschild black hole \cite{Regge:1957td, Zerilli:1970se, Hui:2020xxx, Lenzi:2021wpc, Lenzi:2024tgk}. Although the tensor and vector modes were previously analyzed using the Ishibashi-Kodama formalism \cite{Pereniguez:2021xcj}, we rederive them here within the effective-action framework to ensure a unified and self-consistent treatment of all perturbations channels, which is also necessary for cross-checking known results. Indeed, our results for Love numbers in tensor and vector sectors agree with previous findings in \cite{Pereniguez:2021xcj}.

When restricted to $D=4$, our master equations match with results in \cite{Moncrief:1974gw, Moncrief:1974ng, Chandrasekhar:1985kt}. In $D> 4$, the metric perturbations contain also an intrinsic tensor mode that propagates on its own without mixing with others. This constitutes the simplest perturbation channel. The other components of the metric perturbations inevitably mix with the electromagnetic counterparts.  Due to the different methods used in diagonalizing the perturbations, our master equations do not look the same as those in  \cite{Pereniguez:2021xcj}. However,
the Love numbers in the tensor and vector sectors do agree with those obtained in  \cite{Pereniguez:2021xcj}.
In particular, we confirm that all the Love number vanishes for $4D$ RN black holes \cite{Pereniguez:2021xcj, Rai:2024lho}, a feature similar to the Schwarzschild black hole \cite{Binnington:2009bb,LeTiec:2020bos,Hui:2020xxx}. In $D>4$, we also recover earlier results for the tensor and vector Love numbers \cite{Pereniguez:2021xcj}. So far, the scalar type Love numbers of RN black hole have not been studied.   In $D$-dimensional spacetimes, the effective multipolar index $\tilde{\ell}$ is related to the original multipolar index $\ell$ by $\tilde{\ell} = \ell / (D - 3) $.
We recall that in the case of Schwarzschild black hole,  when $\tilde{\ell}$ is integer valued, all scalar type Love numbers vanish. Hence, it should be interesting to see if similar features appear also in  RN black hole. Our results show that this is indeed the case, which may imply the existence of certain accidental symmetry, as was found for other cases, including $4D$ Schwarzschild black hole \cite{Hui:2021vcv,Charalambous:2021kcz,Parra-Martinez:2025bcu} and RN black hole \cite{Rai:2024lho} and
certain higher dimensional black branes \cite{Charalambous:2025ekl} or black holes \cite{Berens:2025jfs}.

The gauge invariance of the Love number has been addressed in \cite{Hui:2020xxx} by using the worldline effective field theory (WEFT) formalism in which the Love numbers correspond to gauge invariant Wilsonian coefficients in front of higher derivative operators.  Furthermore, it was shown in \cite{Hui:2020xxx} that for Schwarzschild black hole, the Love numbers defined in WEFT match with those extracted directly from the large distance expansion of the perturbative solutions.  Direct computation shows that that the graviton-photon mixing decays sufficiently fast  near the spatial infinity. Consequently, the relation between the WEFT coefficients and the falloff coefficients of perturbations near spatial infinity takes the form as that for Schwarzschild black hole. Thus, in this work we shall extract the Love numbers of RN black holes in all sectors from the large distance expansion of perturbative solutions.

This paper is organized as follows. In section \ref{Quadratic action of perturbations around RN black hole}, we study the full linear perturbations of the RN black hole background in Einstein-Maxwell theory in arbitrary dimensions. We derive the quadratic perturbative quadratic action and use spherical harmonics to decompose it and obtain an effective two-dimensional theory. Note that although our goal is to compute the previously unknown scalar sector, it is necessary to analyse the full linear analysis including both the tensor and vector sectors as well. In Section \eqref{Love numbers in different sectors}, we further simplify the two-dimensional effective theory by introducing auxiliary fields, thereby isolating the true physical degrees of freedom. We diagonalize the graviton-photon mixing sectors and compute the Love numbers. Having properly analyse both the tensor and vector sectors, we focus our attention on the new scalar sector and compute the corresponding Love numbers in general dimensions. We conclude the paper in section 4. We present the explicit forms of the potentials of the scalar sector in the appendix.

\section{Perturbations about the RN black hole \label{Quadratic action of perturbations around RN black hole}}

We begin with the Einstein-Maxwell theory in general $D$ dimensions
\be
S=\frac{1}{16\pi}\int d^Dx\sqrt{-g}\big(R-\frac{1}{4}F^2\big)\ .
\label{Einstein-Maxwell}
\ee
Next we consider perturbing the metric field $g_{\mu\nu}$ and the Maxwell field $A_{\mu}$ around
a background solution denoted as $\bar{g}_{\m\n}$ and $\bar{A}_{\m}$
\bea
g_{\mu\nu}=\bar{g}_{\mu\nu}+h_{\mu\nu},\qquad A_{\mu}=\bar{A}_{\mu}+a_{\mu}\ ,
\label{perturbed}
\eea
where, \(\bar{g}_{\mu\nu}\) and \(\bar{A}_{\mu}\) satisfy the equations of motion
\bea
\bar{R}_{\mu\nu}-\frac{1}{2}\bar{g}_{\mu\nu}\bar{R}-\frac{1}{2}\bar{F}_{\mu\nu}^2+\frac{1}{8}\bar{g}_{\mu\nu}\bar{F}^2=0,\quad
\nabla_\mu \bar{F}^{\mu\nu}=0\ .
\eea
For later convenience, we introduce the following notations
\be
h=\bar{g}^{\mu\nu}h_{\mu\nu},\quad
\bar{F}_{\mu\nu}=\nabla_\mu \bar{A}_\nu-\nabla_\nu \bar{A}_\mu,\quad f_{\mu\nu}=\nabla_\mu a_\nu-\nabla_\nu a_\mu\ ,
\ee
and use $\bar{g}^{\mu\nu}$ and $\bar{g}_{\mu\nu}$ to raise and lower tensorial indices.
Plugging \eqref{perturbed} into \eqref{Einstein-Maxwell} and expand up to quadratic order in perturbations, we obtain
\be
S^{(2)}=\frac{1}{16\pi}\int d^Dx(\mathcal{L}_{a}+\mathcal{L}_{h}+\mathcal{L}_{m})\ ,
\label{quadS}
\ee
where
\bea
\mathcal{L}_{a}&=&-\frac{1}{4}\sqrt{-\bar{g}}f_{\mu\nu}f^{\mu\nu}\ ,
\cr
\mathcal{L}_{h}&=&\sqrt{-\bar{g}}\Big[-\frac{1}{4}\nabla_\lambda h_{\mu\nu}\nabla^\lambda h^{\mu\nu}+\frac{1}{2}\nabla_\lambda h_{\mu\nu}\nabla^\nu h^{\mu\lambda}
-\frac{1}{2}\nabla_\mu h\nabla_\nu h^{\mu\nu}+\frac{1}{4}\nabla_\mu h\nabla^\mu h\Big]\ ,
\cr
\mathcal{L}_{m}&=&\sqrt{-\bar{g}}\Big[f_{\mu\nu}\bar{F}^{\mu\lambda}h^\nu_{\ \lambda}
-\frac{1}{4}\bar{F}^{\mu\nu}\bar{F}^{\rho\sigma}h_{\mu\rho}h_{\nu\sigma}-\frac{1}{4}f_{\mu\nu}\bar{F}^{\mu\nu}h
-\frac{\bar{F}^2}{8(D-2)}h_{\mu\nu}h^{\mu\nu}\nn\\
&&\qquad +\frac{\bar{F}^2}{16(D-2)}h^2\Big]\ .
\label{quadS2}
\eea
It is important to note that the expression for $\mathcal{L}_h$ differs from the given in \cite{Hui:2020xxx,Charalambous:2024tdj} by an overall factor of 2. While this factor is not significant in the case of pure gravity, it becomes crucial in our case, when the graviton-photon mixing occurs.

While \eqref{quadS} holds for any perturbations about any background satisfying \eqref{Einstein-Maxwell}, here we choose the background to be
the $D$-dimensions RN black hole
\bea
d\bar{s}_D^2&=&-fdt^2+\frac{dr^2}{f}+r^2d\Omega^2_{D-2}\,,\qquad f=1-\frac{2\mu}{r^{D-3}}+\frac{q^2}{r^{2(D-3)}}\ ,
\cr
\bar{A}_{(1)}&=&\psi dt\,,\qquad \psi=\sqrt{\frac{2(D-2)}{D-3}}\frac{q}{r^{D-3}}\ .
\label{RN-D}
\eea
Following the procedure of \cite{Hui:2020xxx}, we expand the perturbations in terms of spherical harmonics on $S^{D-2}$ and impose gauge conditions to remove the unphysical redundancies.


Specifically, we denote the line element on the $(D-2)$-dimensional sphere
\be
d\Omega^2_{D-2}=\Omega_{AB}d\theta^Ad\theta^B\ ,
\ee
where $\O_{AB}$ is the metric and the capital Latin letters $ A, B,\cdots=1,2,\cdots D-2$ denote tensor indices on the sphere. The corresponding covariant derivative, \(D_A\), satisfies the condition \(D_A \Omega_{BC} = 0\). We the arrange different components of the perturbations $h_{\mu\nu}, a_{\mu}\,$ into three distinct categories according to their transformation properties under $SO(D-1)$
\begin{itemize}
\item[1.] \(SO(D-1)\) scalars: $a_t$, $a_r$, $h_{tt}$, $h_{tr}$, $h_{rr}$
\item[2.] \(SO(D-1)\) vectors: $a_A$, $h_{tA}$, $h_{rA}$
\item[3.]    \(SO(D-1)\) tensors: $h_{AB}$
\end{itemize}
where the $\m,\n$ indices have been divided into $t, r, A$.
Thus, we need scalar, vector and tensor spherical harmonics in the harmonic expansion. Following \cite{Hui:2020xxx}, for notational simplicity, we omit the summation symbols over \(\ell\), \(m\), and write the expansion as
\bea
a_t(t,r,\theta^A)&=&a_t(t,r)Y_{\ell m}\ ,
\cr
a_r(t,r,\theta^A)&=&a_r(t,r)Y_{\ell m}\ ,
\cr
a_A(t,r,\theta^A)&=&a_{L}(t,r)D_AY_{\ell m}+a_{T}(t,r)Y_{A,\ell m}^{(T)}\ ,
\cr
h_{tt}(t,r,\theta^A)&=&fH_0(t,r)Y_{\ell m}\ ,
\cr
h_{tr}(t,r,\theta^A)&=&H_1(t,r)Y_{\ell m}\ ,
\cr
h_{rr}(t,r,\theta^A)&=&f^{-1}H_2(t,r)Y_{\ell m}\ ,
\cr
h_{tA}(t,r,\theta^A)&=&\mathcal{H}_0(t,r)D_AY_{\ell m}+h_0(t,r)Y_{A,\ell m}^{(T)},\cr
h_{rA}(t,r,\theta^A)&=&\mathcal{H}_1(t,r)D_AY_{\ell m}+h_1(t,r)Y_{A,\ell m}^{(T)}\ ,
\cr
h_{AB}(t,r,\theta^A)&=&r^2\Big[\mathcal{K}(t,r)\Omega_{AB}Y_{\ell m}+G(t,r)D_{\{A} D_{B\}}Y_{\ell m}
\cr
&&+h_2(t,r)D_{(A}Y_{B),\ell m}^{(T)}+h_T(t,r)Y_{AB,\ell m}^{(TT)}
\Big]\ .
\label{mode decomposition}
\eea
Here $Y_{\ell m}$ is scalar harmonic function on $S^{D-2}$, $Y_{A,\ell m}^{(T)}$ denotes the transverse vector harmonics obeying $D^AY_{A,\ell m}^{(T)}=0$, and fianlly $Y_{AB,\ell m}^{(TT)}$ represents the symmetric transverse, traceless rank-2 tensor harmonics satisfying $Y_{AB,\ell m}^{(TT)}=Y_{BA,\ell m}^{(TT)}$, $D^AY_{AB,\ell m}^{(TT)}=0$, $\Omega^{AB}Y_{AB,\ell m}^{(TT)}=0$. The notation $\{AB\}$ means ``symmetric and traceless".

By employing diffeomorphism and U(1) gauge transformations, we make the gauge choice
\be
a_{L}=0,\quad h_2=0,\quad \mathcal{H}_0=0,\quad \mathcal{K}=0,\quad G=0\ .
\ee

When the background metric is given by RN black hole \eqref{RN-D}, the volume element factorizes
\be
\int d^Dx\sqrt{-\bar{g}}=\int dtdr\sqrt{-g_2}\int d^{D-2}\theta\sqrt{\det{\Omega_{AB}}},\quad \sqrt{-g_2}=r^{D-2}\ .
\ee
After substituting the mode decomposition \eqref{mode decomposition} into the quadratic action for perturbations \eqref{quadS}, we perform the integration over $S^{D-2}$ and obtain an effective 2-dimensional action
\be
S^{(2)}=\frac{1}{16\pi}\int dtdr (\widetilde{\mathcal{L}}_a+\widetilde{\mathcal{L}}_h+\widetilde{\mathcal{L}}_m)\ ,
\ee
where we have used the orthonormality conditions satisfied by various spherical harmonics.

Note that the scalar and vector spherical harmonics are both present in the expansion of the U(1) gauge field. Due to their mutual orthogonality, the corresponding perturbations decouple from each other, resulting in the effective action $\widetilde{\mathcal{L}}_a$ of the form
\bea
\widetilde{\mathcal{L}}_a&=&\widetilde{\mathcal{L}}_a^{(s=1)}+\widetilde{\mathcal{L}}_a^{(s=0)}\ ,
\nn\\
\widetilde{\mathcal{L}}_a^{(s=1)}&=&r^{D-4}\left[\frac{1}{2f}\dot{a}_{T}^2-\frac{f}{2}a_{T}'^{2}
-\frac{(\ell+1)(\ell+D-4)}{2r^2}a_{T}^2
\right]\ ,
\nn\\
\widetilde{\mathcal{L}}_a^{(s=0)}&=&r^{D-4}\left[\frac{r^2}{2}(\dot{a}_r-a_t')^2+\frac{\ell(\ell+D-3)}{2f}a_t^2-\frac{\ell(\ell+D-3)}{2}fa_r^2
\right]\ .
\label{MWaction}
\eea
Since ${\cal L}_a$ in \eqref{quadS2} does not contain any term involving the background electric field, the above results agree with those found in \cite{Hui:2020xxx} for a U(1) gauge field perturbing around an arbitrary spherically symmetric black hole.

The action $\widetilde{\mathcal{L}}_h$ receives contributions from scalar, vector, and tensor modes, which do not mix with each other. We thus have
\bea
\widetilde{\mathcal{L}}_h=\widetilde{\mathcal{L}}_h^{(s=2)}+\widetilde{\mathcal{L}}_h^{(s=1)}+\widetilde{\mathcal{L}}_h^{(s=0)}\ ,
\eea
where the effective action in each sector takes the form below, after proper integration by parts is done
\bea
&&\widetilde{\mathcal{L}}_h^{(s=2)}=\frac{r^{D-2}}{2}\Bigg[\frac{1}{2f}\dot{h}_T^2-\frac{f}{2}h_T'^2+f\Big(
\frac{D-3}{r^2}+\frac{6-2D-\ell(\ell+D-3)}{2fr^2}+\frac{f'}{rf}
\Big)h_T^2
\Bigg]\ ,\nn\\
&&\widetilde{\mathcal{L}}_h^{(s=1)}=\frac{r^{D-4}}{2}\Bigg[
\Big(\dot{h}_1+\frac{2}{r}h_0-h_0'\Big)^2+\frac{2(D-3)f+2rf'-(\ell+1)(\ell+D-4)}{r^2}fh_1^2
\nn\\
&&\qquad\qquad\qquad +\frac{(\ell+1)(\ell+D-4)-2(D-3)f-2rf'}{r^2f}h_0^2
\Bigg]\ ,\nn\\
&&\widetilde{\mathcal{L}}_h^{(s=0)}=\frac{\ell(\ell+D-3)r^{D-4}}{2}\Bigg[
\dot{\mathcal{H}}_1^2+\frac{2f}{r^2}\big((D-3)f+rf'\big)\mathcal{H}_1^2
+\frac{(D-2)\big((D-3)f+rf'\big)}{2\ell(\ell+D-3)}H_2^2
\nn\\
&&\qquad\quad  +H_0\Bigg(\big(f'+\frac{2(D-3)f}{r}\big)\mathcal{H}_1+2f\mathcal{H}_1'-\Big(
1+\frac{(D-2)\big((D-3)f+rf'\big)}{\ell(\ell+D-3)}\Big)H_2
\nn\\
&&\qquad\quad  -\frac{(D-2)rfH_2'}{\ell(\ell+D-3)}
\Bigg)-\frac{2(D-3)f+rf'}{r}H_2\mathcal{H}_1
+H_1\Big(H_1-2\dot{\mathcal{H}}_1+\frac{2(D-2)r}{\ell(\ell+D-3)}\dot{H}_2
\Big)
\Bigg]
\nn\\
&&\qquad\quad +\frac{r^{D-3}}{8}\Big(rf''+(D-2)f'\Big)\Big((H_0+H_2)^2-4H_1^2\Big)\ .
\label{FP action}
\eea
It is worth noting that, compared to the result in \cite{Hui:2020xxx}, our expression for $\widetilde{\mathcal{L}}_h^{(s=0)}$ includes an additional term in the last line. This is because \cite{Hui:2020xxx} used the background field equations to simplify the result. Moreover, for Schwarzschild black hole, this term vanishes. However, for RN black hole, this term does not vanish due to the presence of electric charge.

Finally, we present the explicit forms of
$\widetilde{\mathcal{L}}_m$. According to the representations of perturbations under $SO(D-1)$, $\widetilde{{\cal L}}_m$ is separated into three pieces
\be
\widetilde{\mathcal{L}}_m=\widetilde{\mathcal{L}}_m^{(s=2)}+\widetilde{\mathcal{L}}_m^{(s=1)}+\widetilde{\mathcal{L}}_m^{(s=0)}\ ,
\ee
where
\bea
&&\widetilde{\mathcal{L}}_m^{(s=2)}=\frac{r^{D-2}\psi'^2}{4(D-2)}h_T^2,
\nn\\
&&\widetilde{\mathcal{L}}_m^{(s=1)}=r^{D-4}\psi'\big(\dot{a}_{T}h_1-a_{T}'h_0\big)
+\frac{r^{D-4}\psi'^2}{2(D-2)}\big(fh_1^2-f^{-1}h_0^2
\big),
\nn\\
&&\widetilde{\mathcal{L}}_m^{(s=0)}=\frac{r^{D-4}\psi'}{2}\Big[r^2(\dot{a}_r-a_t')(H_2-H_0)-2\ell(\ell+D-3)a_t\mathcal{H}_1
\Big]\nn\\
&&+\frac{r^{D-4}\psi'^2}{8(D-2)}\Big[4\ell(\ell+D-3)f\mathcal{H}_1^2+r^2\big(
H_0^2+H_2^2+4(D-3)H_1^2-2(2D-5)H_0H_2
\big)
\Big]\ .
\label{linear mix}
\eea
\section{Love numbers in different sectors \label{Love numbers in different sectors}}
We now proceed to solve for the perturbations in the static limit, from which we read off the Love numbers.

\subsection{Tensor Love numbers in diverse dimensions}\label{Tensor Love numbers}
Let us first focus on the tensor sector which is the simplest one to study. By ``tensor", we mean the corresponding perturbations are expanded in terms of rank-2 symmetric traceless harmonics on $S^{D-2}$.
Compared with the tensor sector action for Schwarzschild black hole \cite{Hui:2020xxx,Charalambous:2024tdj}, the tensor mode action obtained here receives contributions from nontrivial background electric field. Specifically, the action takes the form
\bea
&&\mathcal{L}^{(s=2)}=\widetilde{\mathcal{L}}_h^{(s=2)}+\widetilde{\mathcal{L}}_m^{(s=2)}
\nn\\
&&=\frac{r^{D-2}}{2}\Bigg[\frac{1}{2f}\dot{h}_T^2-\frac{f}{2}h_T'^2+f\Big(
\frac{D-3}{r^2}+\frac{6-2D-\ell(\ell+D-3)}{2fr^2}+\frac{f'}{rf}+\frac{\psi'^2}{2(D-2)f}
\Big)h_T^2
\Bigg],
\label{tensor action hT}
\eea
where the last term comes from the energy momentum tensor of the background electric field, and thus vanishes in the case of Schwarzschild black hole.
To rewrite the action in standard form, we redefine the field \(\Psi_T\)
\be
\Psi_T(t,r)=\frac{1}{\sqrt{2}}r^{\frac{D-2}{2}}h_T\ ,
\ee
and perform the necessary integration by parts which leads to
\bea
\mathcal{L}^{(s=2)}&=&\frac{1}{2f}\dot{\Psi}_T^2-\frac{f}{2}\Psi_T'^2-\frac{1}{2}V_T(r)\Psi_T^2\ ,\nn\\
V_T(r)&=&\frac{\ell(\ell+D-3)+2(D-3)}{r^2}+\frac{D^2-14D+32}{4r^2}f+\frac{D-6}{2r}f'-\frac{\psi'^2}{D-2}\ .
\label{tensor perturbation action}
\eea

As demonstrated in \cite{Hui:2020xxx}, the tensor Love number is determined from the transverse-traceless component of \( h_{\mu\nu} \) on the $S^{D-2}$ direction. In particular, $\Psi_T$ is related to the gauge-invariant tensor $h_{AB}^{TT}$ by
\be
h_{AB}^{TT}=\sqrt2 r^{\frac{2-D}{2}}\Psi_T Y_{AB,\ell m}^{(TT)}\ .
\ee
After performing separation of variables for $\Psi_T$ as
\be
\Psi_T(t,r)=\frac{1}{\sqrt{2}}r^{\frac{D-2}{2}}e^{-i\omega t}R_\ell(r)\ ,
\ee
and adopting new variables below
\be
x=\frac{r^{D-3}-\tilde{r}_+}{\tilde{r}_+-\tilde{r}_-},\quad  \tilde{r}_\pm=r_\pm^{D-3}=\mu\pm\sqrt{\mu^2-q^2},   \quad \tilde{\ell}=\frac{\ell}{D-3}\ ,
\label{coord x}
\ee
we obtain equation satisfied by $R_\ell(r)$
\be
x(1+x)R_\ell''(x)+(2x+1)R_\ell'(x)-\Big[\tilde{\ell}(\tilde{\ell}+1)-\frac{\big(\tilde{r}_+(1+x)-\tilde{r}_-x\big)^{\frac{2(D-2)}{D-3}}}{(D-3)^2(\tilde{r}_+-\tilde{r}_-)^2x(1+x)}
\omega^2\Big] R_\ell(x)=0\ .
\label{radial EOM tensor}
\ee
In the static limit \(\omega = 0\), the equation above reduces to that of a scalar perturbation in the background of Schwarzschild black hole \cite{Hui:2020xxx,Charalambous:2021mea}. Similarly, the solution that remains regular on the horizon \(x = 0\) is given by
\bea
R_\ell= \, _2F_1(\tilde{\ell}+1,-\tilde{\ell};1;-x)\ .\label{Rl}
\eea
Not surprisingly, this tensor Love numbers take similar form as those of a scalar perturbation in the background of Schwarzschild black hole \cite{Hui:2020xxx}.

To compare with \cite{Pereniguez:2021xcj}, we can also perform a coordinate transformation $x \rightarrow z =\left(\frac{r_+}{r} \right)^{D-3}$ on (\ref{radial EOM tensor}) to convert it into the standard Fuchsian form
\bea
R_{\ell}''(z)+\frac{P(z) }{z}R_{\ell}'(z)+\frac{Q(z)}{z^2}R_{\ell}(z)=0\ .
\label{EOM tensor z}
\eea
The black hole horizon and spatial infinity reside at $z=1$ and $z=0$ respectively.
The indices at  singular point $z=0$ are $\tilde{\ell}+1$ and $-\tilde{\ell}$. Therefore, we consider the following cases separately.
\begin{enumerate}
    \item $2\tilde{\ell}+1 \in \mathbb{N}$: In a neighborhood of $z=0$, the general solution has the form
\bea
R_{\ell}(z) = A z^{\tilde{\ell}+1} \Phi_{\rm resp}(z)+ B \left(z^{-\tilde{\ell}}\Phi_{\rm tidal}(z) + R z^{\tilde{\ell}+1} \Phi_{\rm resp}(z)\ln z \right)\ .
\eea
In this way of parametrization, $\Phi_{\rm resp}$ and $\Phi_{\rm tidal}$ are analytic functions in a neighborhood of $z = 0$. The ratio between $A$ and $B$ is fixed, once demanding regularity of the solution at the horizon. Hence the solution takes the form
\bea
R_{\ell}(z) = Bz^{\tilde{\ell}+1} \left( k  \Phi_{\rm resp}(z)+ z^{-2\tilde{\ell}-1}\Phi_{\rm tidal}(z) + R \Phi_{\rm resp}(z)\ln z \right)\ .
\label{RT}
\eea
As discussed in \cite{Pereniguez:2021xcj},  the presence of the logarithmic term implies that the exact value of $k$ is ambiguous. However, the precise value of $R$ can be obtained by substituting $z^{-\tilde{\ell}}\Phi_{\rm tidal}+R z^{\tilde{\ell}+1}\Phi_{\rm resp}\ln{z}$ into (\ref{EOM tensor z}) and solve the equation order by order in small $z$ expansion \cite{Pereniguez:2021xcj}. In terms of the radial coordinate $z$, we have
\bea
R=\frac{(-1)^{2 \tilde{\ell}} \Gamma (\tilde{\ell}+1)^2}{(2 \tilde{\ell})! (2 \tilde{\ell}+1)! \Gamma (-\tilde{\ell})^2}(1-\sigma )^{2 \tilde{\ell}+1},\quad \sigma:=\tilde{r}_-/\tilde{r}_+\ .
\eea
Here and below,the effect of electric charge is encoded in the parameter $\sigma$.
When $\tilde{\ell}$ is  an integer, the coefficient $R$ vanish. Meanwhile, the solution regular at the horizon consists of only $\Phi_{\rm tidal}$ which is a polynomial of degree $<2\tilde{\ell}+1$. Therefore, when $\tilde{\ell}$ is an integer, tensor Love number always vanishes.

\item $2\tilde{\ell}+1 \notin \mathbb{N}$: In this case, from the asymptotic expansion of the solution (\ref{Rl}) at infinity,  we can read off the tensor Love number directly,
 \bea
k_T=\frac{\Gamma(-2\tilde{\ell}-1)\Gamma(\tilde{\ell}+1)^2}{\Gamma(2\tilde{\ell}+1)\Gamma(-\tilde{\ell})^2}(1-\sigma)^{2\tilde{\ell}+1} \quad\text{(in terms of $z$)}\ ,
\eea
which allows us to test the numerical approach, which will later be applied to the vector and scalar sectors in which  analytical expressions of the Love numbers are currently beyond the scope. After imposing regularity at the horizon, the general solution of Eq.(\ref{EOM tensor z}) takes the form
\bea
R_{\ell}(z) = B\left( k_T z^{\tilde{\ell}+1} \Phi_{\rm resp}(z)+ z^{-\tilde{\ell}}\Phi_{\rm tidal}(z)\right).\label{general}
\eea
The numerical value of $k_T$ is determined by matching the solutions near the horizon $(z=1)$ and at infinity $(z=0)$. The regularity at the horizon serves as the key physical boundary condition. This requirement dictates that only the regular solution branch is selected near the horizon. By performing a series expansion in this region, the branch with $\log(1-z)$ is discarded. The regular solution at $z=1$ is solved as a series expansion in powers of $(z-1)$ up to order $(z-1)^{15}$. We then evaluate the series and its derivative at $z=1-10^{-8}$ which are used as initial values in the numerical integration of the differential equation towards $z=0$. The resulting numerical solution is matched with the asymptotic behavior near $z=0$ \eqref{general} to determine the value of $k_T$. In Fig.\ref{kT}, we plot both the analytical (solid line) and numerical (points) results for the tensor Love numbers in several cases obeying $2\tilde{\ell}+1 \notin \mathbb{N}$. The excellent agreement between the numerical and theoretical values demonstrates the validity of our method.
\begin{figure}
    \centering
    \includegraphics[width=0.48\linewidth]{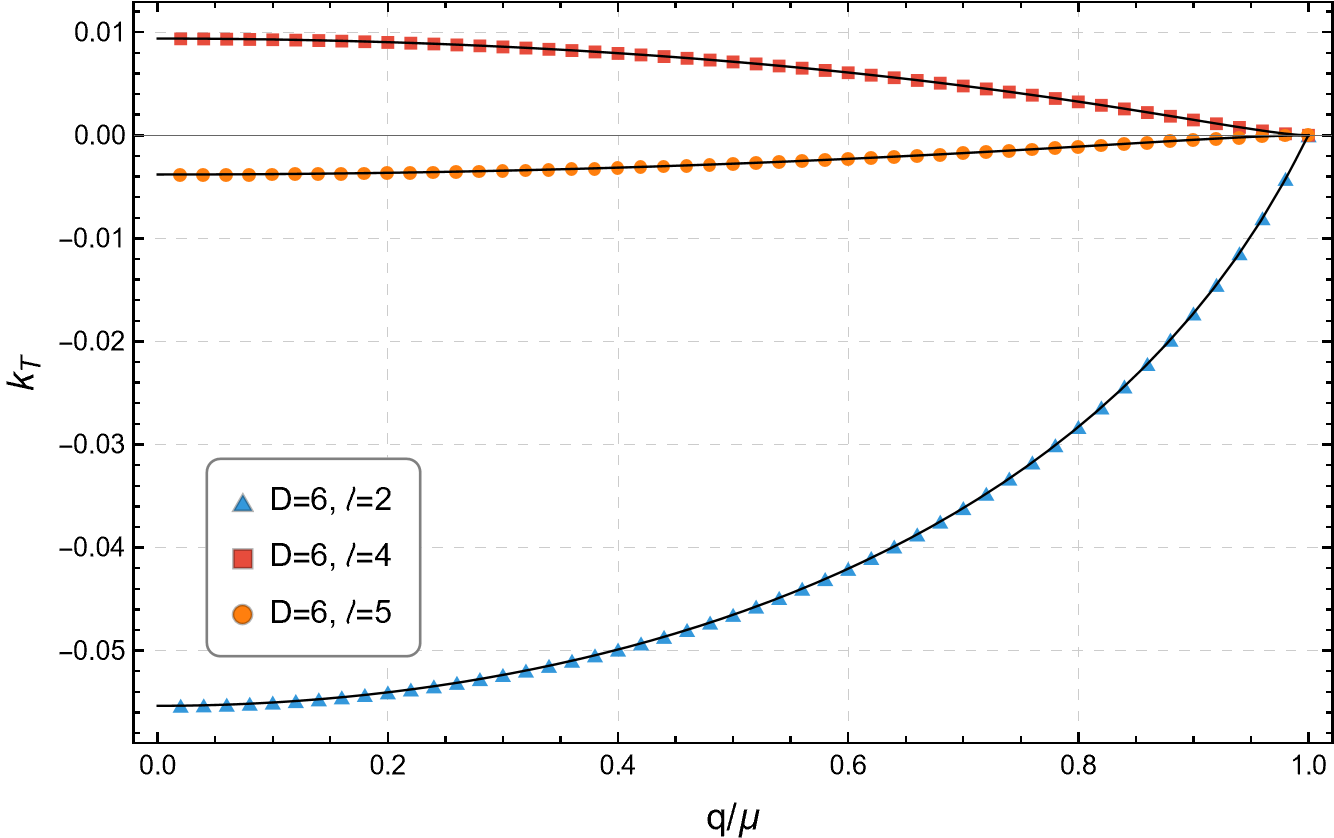}
    \includegraphics[width=0.48\linewidth]{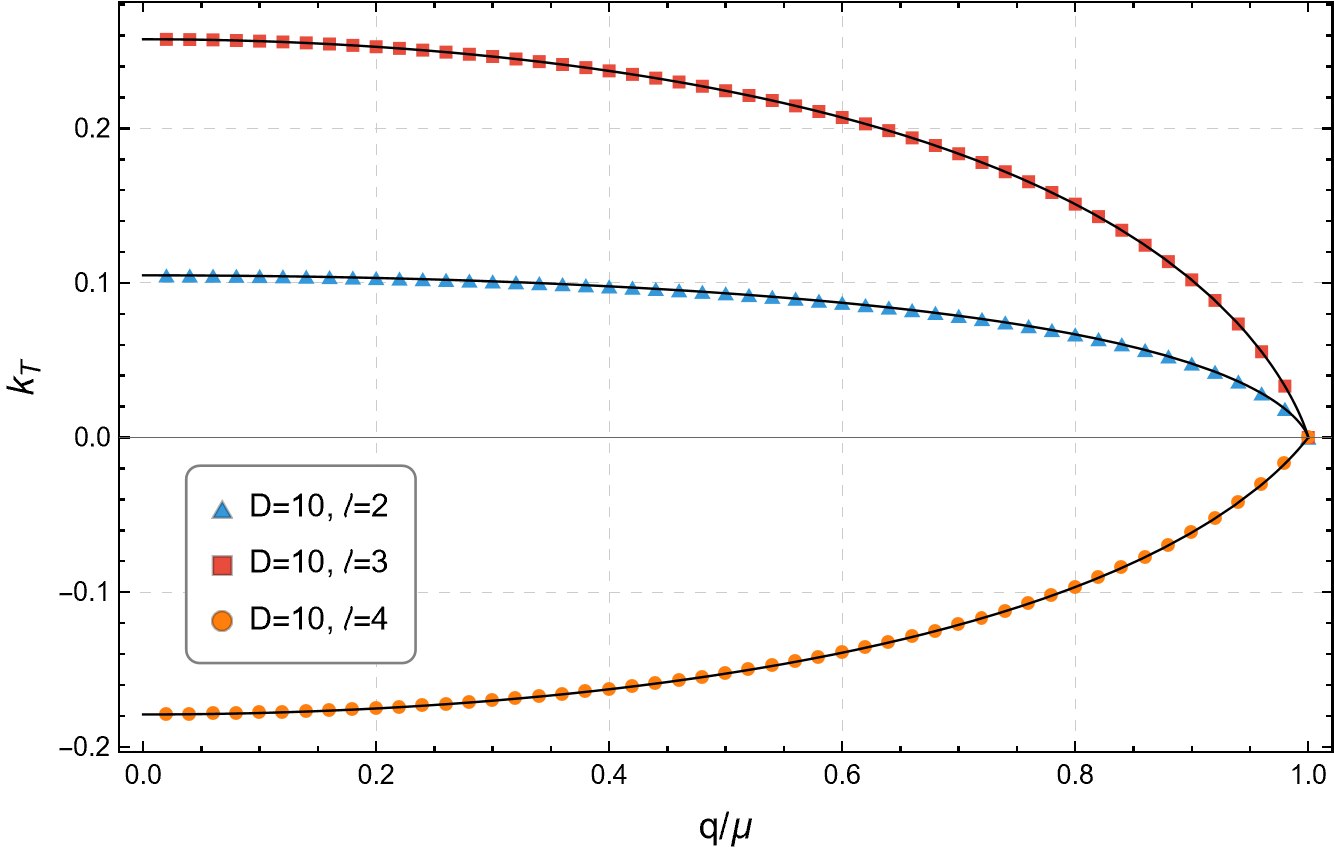}
    \caption{These plots show in $D=6,10$, and $2\tilde{\ell}+1\notin \mathbb{N}$, the dependence of tensor Love numbers $k_T$ on $q/\mu$. Solid lines correspond to analytical values and numerical values are denoted by different markers.}
    \label{kT}
\end{figure}

\end{enumerate}

It is important to note that in $D = 4$, the intrinsic tensor harmonics on $S^2$ is absent \cite{Binnington:2009bb}. Consequently, in $D=4$, the tidal deformations of an RN black hole are characterized by the gravito-electric response and the gravito-magnetic response coefficients, which are associated with the scalar and vector-type perturbations respectively. It is well known that for Schwarzschild black hole, both of these coefficients vanish in $D=4$ \cite{Hui:2020xxx}. Later, we will compute gravito-electric response and the gravito-magnetic response coefficients for RN black holes in dimensions $D\ge4$.

\subsection{Effective action of vector type perturbations}

Unlike the tensor sector \eqref{tensor action hT}, the Regge-Wheeler mode in the gravitational perturbation and the magnetic mode in the Maxwell field are both vector-type perturbations, and thus can mix with each other. As a result, the action in the vector sector is more involved than that in the tensor sector \eqref{tensor action hT}. In this section, we will show how to diagonalize the vector-type perturbations by introducing an auxiliary field.

Since both the gravitational and Maxwell field perturbations contain vector modes, they are coupled to each other. As a result, the effective Lagrangian in this sector is of the form
\be
\mathcal{L}^{(s=1)}=\widetilde{\mathcal{L}}_h^{(s=1)}+\widetilde{\mathcal{L}}_a^{(s=1)}+\widetilde{\mathcal{L}}_m^{(s=1)}.\label{vector sector}
\ee
After performing the appropriate integration by parts, the action for the vector sector can be written as
\bea
\mathcal{L}^{(s=1)}&=&r^{D-4}\Bigg[
\frac{1}{2}\Big(\dot{h}_1+\frac{2}{r}h_0-h_0'-a_{T}\psi'\Big)^2+\Big(\psi''+\frac{D-2}{r}\psi'\Big)a_{T}h_0
\nn\\
&&+(fh_1^2-f^{-1}h_0^2)\Big(
\frac{(D-3)f+rf'}{r^2}+\frac{\psi'^2}{2(D-2)}-\frac{(\ell+1)(\ell+D-4)}{2r^2}
\Big)
\nn\\
&&+\frac{1}{2f}\dot{a}_{T}^2-\frac{f}{2}a_{T}'^{2}
-\frac{1}{2}\Big(\frac{(\ell+1)(\ell+D-4)}{r^2}+\psi'^2\Big)a_{T}^2
\Bigg]\ .
\label{s=1}
\eea

To diagonalize all the perturbations,   we introduce an auxiliary field \(Q_{\mathrm{aux}}(t,r)\) as in \cite{Hui:2020xxx}, and reexpress the action \eqref{s=1} as
\bea
\mathcal{L}^{(s=1)}_{\mathrm{aux}}&=&r^{D-4}\Bigg[
Q_{\mathrm{aux}}\Big(\dot{h}_1+\frac{2}{r}h_0-h_0'-a_{T}\psi'\Big)-\frac{1}{2}Q_{\mathrm{aux}}^2+\Big(\psi''+\frac{D-2}{r}\psi'\Big)a_{T}h_0
\nn\\
&&+(fh_1^2-f^{-1}h_0^2)\Big(
\frac{(D-3)f+rf'}{r^2}+\frac{\psi'^2}{2(D-2)}-\frac{(\ell+1)(\ell+D-4)}{2r^2}
\Big)
\nn\\
&&+\frac{1}{2f}\dot{a}_{T}^2-\frac{f}{2}a_{T}'^{2}
-\frac{1}{2}\Big(\frac{(\ell+1)(\ell+D-4)}{r^2}+\psi'^2\Big)a_{T}^2
\Bigg]\ .
\label{s=1 aux}
\eea
Upon solving for \(Q_{\mathrm{aux}}\) from its equation of motion and substituting back, the action above \eqref{s=1 aux} recovers the original action \eqref{s=1} without the auxiliary field.  On the other hand, one can first solve  for \(h_0\) and \(h_1\) in terms of \(Q_{\mathrm{aux}}\) as
\be
h_0=-\frac{rf}{(\ell-1)(\ell+D-2)}\big[(D-2)Q_{\mathrm{aux}}+rQ_{\mathrm{aux}}'\big]\ ,\quad
h_1=-\frac{r^2}{(\ell-1)(\ell+D-2)f}\dot{Q}_{\mathrm{aux}}\ .
\ee
Here, the expressions above have been simplified using the fact that the background RN black hole solution \eqref{RN-D} satisfies
\bea
f''+\frac{D-2}{r}f'-\frac{D-3}{D-2}\psi'^2=0,\quad \psi''+\frac{D-2}{r}\psi'=0\ .
\eea
Therefore, the vector mode from the metric has only one independent degree of freedom. Next, we introduce \(\Psi_{\mathrm{RW}}\) and \(\Psi_V\) below
\be
\Psi_{\mathrm{RW}}=\sqrt{\frac{r^{D-2}}{(\ell-1)(\ell+D-2)}}Q_{\mathrm{aux}},\quad \Psi_V=\sqrt{(\ell-1)(\ell+D-2)r^{D-4}}a_{T}\ ,
\ee
in terms of which the action \eqref{s=1 aux} can be transformed into
\bea
\mathcal{L}^{(s=1)}&=&\frac{1}{2f}\dot{\Psi}_{\mathrm{RW}}^2-\frac{f}{2}\Psi_{\mathrm{RW}}'^2-\frac{1}{2}V_{\mathrm{RW}}(r)\Psi_{\mathrm{RW}}^2
\nn\\
&&+\frac{1}{(\ell-1)(\ell+D-2)}\Big(\frac{1}{2f}\dot{\Psi}_V^2-\frac{f}{2}\Psi_V'^2-\frac{1}{2}V_V(r)\Psi_V^2\Big)-\frac{\psi'}{r}\Psi_{\mathrm{RW}}\Psi_V\ ,
\nn\\
V_{\mathrm{RW}}(r)&=&\frac{(\ell+1)(\ell+D-4)}{r^2}+f\frac{(D-6)(D-4)}{4r^2}-f'\frac{D+2}{2r}-\frac{\psi'^2}{D-2}\ ,
\nn\\
V_V(r)&=&\frac{(\ell+1)(\ell+D-4)}{r^2}+f\frac{(D-6)(D-4)}{4r^2}+f'\frac{D-4}{2r}+\psi'^2\ .
\label{action RW V}
\eea
From the expression of the action, we see that the Regge-Wheeler mode is coupled to the magnetic vector mode due to the nontrivial background electric field. Specific to \(D = 4\), we find that the equations of motion for \(\Psi_{\mathrm{RW}}\) and \(\Psi_V\) match with those previous obtained in \cite{Moncrief:1974gw}\footnote{Here, \(\Psi_{\mathrm{RW}}\) differs from \(\pi_g\) in \cite{Moncrief:1974gw} by an overall constant , namely
$
\Psi_{\mathrm{RW}}=\frac{\pi_g}{\ell(\ell-1)(\ell+1)(\ell+2)}$. }

The gauge-invariant vector-type Love numbers are encoded in the magnetic components of the Maxwell field strength and linearized Weyl tensor,
\be
F_{AB}=\partial_Aa_B-\partial_B a_A,\quad C_{trAB}=2r^2\partial_r(r^{-2}D_{[A}h_{B]t})\ .
\ee
Plugging \eqref{mode decomposition}, one obtains \cite{Hui:2020xxx}
\be
F_{AB}=r^{\frac{4-D}{2}}\Psi_VD_{[A}Y_{B],\ell m}^{(T)},\quad C_{trAB}\xrightarrow{r\rightarrow\infty}r^{\frac{2-D}{2}}\Psi_{\mathrm{RW}}D_{[A}Y_{B],\ell m}^{(T)}\ ,
\ee
which can be matched to the results derived using WEFT. The matching procedure has been carried out for Schwarzschild black hole in \cite{Hui:2020xxx}, confirming the agreement between Love numbers extracted from the large distance expansions of $\Psi_V\,,\Psi_{\mathrm{RW}}$ and those defined using wilsonian coefficients on the WEFT. The Love numbers derived from \(\Psi_V\) and \(\Psi_{\mathrm{RW}}\) are also known as the magnetic response coefficient and the gravito-magnetic response coefficient.

To diagonalizing \eqref{action RW V}, we follow the strategy of \cite{Moncrief:1974gw} by first rewriting \(\Psi_{\mathrm{RW}}\) and \(\Psi_V\) as linear combinations of \(\mathcal{W}_{\pm}(t,r)\)
\be
\Psi_{\mathrm{RW}}=\alpha_1\mathcal{W}_{+}+\alpha_2\mathcal{W}_{-},\quad  \Psi_{V}=\alpha_3\mathcal{W}_{+}+\alpha_4\mathcal{W}_{-}\ .
\label{decouple vector}
\ee
By demanding that $\mathcal{W}_{\pm}(t,r)$ decouple from each other, we can solve for the coefficients $\a_i $s which turn out to be constant in this case \footnote{In general, they can be functions of the radial coordinate $r$.}
\begin{align}
\alpha_1&=\Big[1+\frac{c_1^2}{(\ell-1)(\ell+D-2)}
\Big]^{-\frac{1}{2}}\,,  &\alpha_2&=\Big[1+\frac{(\ell-1)(\ell+D-2)}{c_1^2}
\Big]^{-\frac{1}{2}},
\nn\\
\alpha_3&=c_1\alpha_1\,, & \alpha_4&=-\frac{(\ell-1)(\ell+D-2)}{c_1}\alpha_2\ ,
\end{align}
where the constant $c_1$ is given by
\be
c_1=\frac{\sqrt{2} \sqrt{D-2}  (\ell -1) (\ell+D -2)q}{(D-1) \sqrt{D-3} \mu +\sqrt{(D-3) (D-1)^2 \mu ^2+2 (D-2)  (\ell -1) (\ell+D -2)q^2}}\ .
\ee
In terms of \(\mathcal{W}_{\pm}(t,r)\), the Lagrangian \eqref{action RW V} becomes
\bea
&&\mathcal{L}^{(s=1)}=\mathcal{L}_{\mathcal{W}+}+\mathcal{L}_{\mathcal{W}-}\ ,
\nn\\
&&\mathcal{L}_{\mathcal{W}+}=\frac{1}{2f}\dot{\mathcal{W}}_{+}^2-\frac{f}{2}\mathcal{W}_{+}'^2-\frac{1}{2}V_{\mathcal{W}+}(r)\mathcal{W}_{+}^2\ ,
\cr
&&\mathcal{L}_{\mathcal{W}-}=\frac{1}{2f}\dot{\mathcal{W}}_{-}^2-\frac{f}{2}\mathcal{W}_{-}'^2-\frac{1}{2}V_{\mathcal{W}-}(r)\mathcal{W}_{-}^2\ ,
\cr
&&V_{\mathcal{W}+}=\frac{(D-6) (D-4) f+4 (\ell +1) (\ell+D -4)}{4 r^2}
\cr
&&+\frac{1}{2 (D-2)  \big(c_1^2+(\ell -1) (\ell+D -2)\big)r}\Bigg[
(D-2) f' \big(c_1^2 (D-4)-(D+2) (\ell -1) (\ell+D -2)\big)
\cr
&&+4 c_1 (D-2) (\ell -1) (\ell+D -2) \psi '+2 r \big(c_1^2 (D-2)-(\ell -1) (\ell+D -2)\big)\psi'^2
\Bigg]\ ,
\cr
&&V_{\mathcal{W}-}=\frac{(D-6) (D-4) f+4 (\ell +1) (\ell+D -4)}{4 r^2}\nn\\
&&-\frac{1}{2 (D-2)  \big(c_1^2+(\ell -1) (\ell+D -2)\big)r}\Bigg[
(D-2) f' \big[c_1^2 (D+2)- (D-4)(\ell -1) (\ell+D -2)\big]
\cr
&&+4 c_1 (D-2) (\ell -1) (\ell+D -2) \psi '+2 r \big(c_1^2 -(D-2)(\ell -1) (\ell+D -2)\big)\psi'^2
\Bigg]\ .
\label{wlag}
\eea

For Schwarzschild black hole with $q=0$, the combination coefficients in \eqref{decouple vector} reduce to
\bea
\alpha_1=1,\quad \alpha_2=\alpha_3=0,\quad \alpha_4=-\sqrt{(\ell-1)(\ell+D-2)}.\label{decouple vector0}
\eea

\subsection{Vector-type Love numbers in diverse dimensions}\label{Love number in different dimensions}

Now we proceed to solve for $\mathcal{W}_{\pm}(t,r)$. Plugging the ansatz below into the equations of motion from \eqref{wlag}
\be
\mathcal{W}_{\pm}(t,r)=e^{-i\omega t}r^{\frac{D-2}{2}}R_{\mathcal{W}\pm}(r)\ ,
\label{separation RW V}
\ee
we obtain
\bea
&&x(1+x)R_{\mathcal{W}\pm}''+(1+2x)R_{\mathcal{W}\pm}'+\frac{\omega ^2r_+^2 \big((1-\s) x+1\big)^{\frac{2 (D-2)}{D-3}}}{(D-3)^2 (1-\s)^2 x (x+1)}R_{\mathcal{W}\pm}
\nn\\
&&+\Big[\frac{(D^2-4 D+5) (1+\s)\pm c_{\mathcal{W}}}{2 (D-3)^2 \big((1-\s) x+1\big)}
-\frac{(D-2) (2 D-5) \s }{(D-3)^2 \big((1-\s) x+1\big)^2}-\tilde{\ell}  (\tilde{\ell} +1)\Big]R_{\mathcal{W}\pm}=0\ ,
\label{radial EOM RW and V}
\eea
where $\sigma:=\tilde{r}_-/\tilde{r}_+$ (see \eqref{coord x} for the definition of $\tilde{r}_\pm$ ) and the new  constant \(c_\mathcal{W}\) is defined as
\be
c_{\mathcal{W}}= \sqrt{(D-3)^2 (D-1)^2 (1+\sigma)^2
-8 (D-2)(D-3)  \big(D-2-\tilde{\ell } (\tilde{\ell }+1)(D-3)^2 \big) \sigma}\ .
\ee
In the static case $\omega=0$, solutions to the equations above can be expressed in terms of the Heun function. After imposing regular boundary condition at the outer horizon $x=0$, the solution is
\bea
R_{\mathcal{W}\pm}&=&\big((1-\sigma) x+1\big)^{-\frac{D-2}{D-3}}\text{HeunG}\Big[\frac{1}{1-\sigma},q_{\pm},\alpha ,\beta ,1 ,1 ,-x\Big]\ ,
\nn\\
q_{\pm}&=&-\frac{\mp c_{\mathcal{W}}+2  \big((D-3)^2 \tilde{\ell}  (\tilde{\ell} +1)-D+2\big)+(D-3) (D-1) (1+\sigma)}{2 (D-3)^2 (1-\sigma)}\ ,
\nn\\
\alpha&=&-\frac{D-2}{D-3}-\tilde{\ell},\quad \beta=\tilde{\ell} -\frac{1}{D-3}\ .
\label{RW V solution 0}
\eea
Had we known the connection for Heun function, the solution above would then be expressed as a
linear combination of two other solutions with definite and distinct fall-offs near infinity, from which we read off the Love numbers. However, due to the lacking of the connection formula for Heun function, one still needs to employ numerical method.  We thus adopt the same approach as in the previous section which applies to all Fuchsian type equations.

Through a coordinate transformation $x \rightarrow z =\left(\frac{r_+}{r} \right)^{D-3}=1/((1-\sigma)x+1)$, we can convert (\ref{radial EOM RW and V}) into the standard Fuchsian form. For equations of this type, the indices near the singular point $z=0$ are $-\tilde{\ell}$ and $\tilde{\ell}+1$ and the analysis can be divided into the two cases:

\begin{enumerate}
    \item $2\tilde{\ell}+1 \in \mathbb{N}$:
    Similarly, by imposing the regularity condition of the solution at the horizon, we fix the ratio of the two independent solutions, and the solution takes the form near $z=0$
\bea
R_{\mathcal{W}\pm}(z) = B_{\pm} \left(k_\pm z^{\tilde{\ell}+1} \Psi_{\text{resp}\pm}(z) + z^{-\tilde{\ell}}\Psi_{\text{tidal}\pm}(z) + R_{\pm} z^{\tilde{\ell}+1} \Psi_{\text{resp}\pm}(z) \ln z \right)\ ,
\label{Rw}
\eea
where $\Psi_{\rm resp}$ and $\Psi_{\rm tidal}$ are analytic functions in a neighborhood of $z = 0$.

The exact value of $R_{\pm}$ can be obtained using the method described in Sec.\ref{Tensor Love numbers}. As an example, when $\tilde{\ell}=1$, we have
\bea
R_\pm=&&-\frac{1}{96 (D-3)^6}\Big[\left(D^2 (\sigma +1)-8 D (\sigma +1)\mp c_{\mathcal{W}}+13 (\sigma +1)\right)\cr
&&\big(D^4 \left(\sigma ^2+34 \sigma +1\right) -12 D^3 \left(\sigma ^2+30 \sigma +1\right)+2 D^2 \left(25 \sigma ^2+706 \sigma +25\right)
\cr
&&-4 D \left( \pm c_{\mathcal{W}} \sigma \pm c_{\mathcal{W}}+23 \sigma ^2+610 \sigma +23\right)- c_{\mathcal{W}}^2\cr
&&\pm 8  c_{\mathcal{W}} (\sigma +1)+65 \sigma ^2+1570 \sigma +65\big)\Big],
\eea
It can be seen that $\Psi_{\text{resp}}$ can be completely absorbed order by order into $\Psi_{\text{tidal}}$, yielding
\bea
R_{\mathcal{W}\pm}(z) = B_{\pm}\left(z^{-\tilde{\ell}}\Psi'_{\text{tidal}\pm}(z) + R_{\pm} z^{\tilde{\ell}+1} \Psi_{\text{resp}\pm}(z) \ln z\right).
\eea

Near spatial infinity ($z \rightarrow 0$)  , the asymptotic behaviors of $\Psi_{\mathrm{RW}}$ and  $\Psi_{V}$ is then
\bea
r^{\frac{2-D}{2}}\Psi_{\mathrm{RW}}|_{z\rightarrow 0} &= z^{\tilde{\ell}+1}\left(\alpha_1 B_{+} R_{+} + \alpha_2 B_{-} R_{-}\right) \ln z + z^{-\tilde{\ell}}(\alpha_1 B_{+} + \alpha_2 B_{-})\ , \\
r^{\frac{2-D}{2}}\Psi_{V}|_{z\rightarrow 0} &= z^{\tilde{\ell}+1}\left(\alpha_3 B_{+} R_{+} + \alpha_4 B_{-} R_{-}\right) \ln z + z^{-\tilde{\ell}}(\alpha_3 B_{+} + \alpha_4 B_{-})\ .
\eea
Therefore, the Love numbers are defined as (coefficient of the logarithmic term over coefficient of the source term):
\be
k_{\mathrm{RW}}=\frac{\alpha_1 B_{+}R_{+}+\alpha_2 B_{-}R_{-}}{\alpha_1B_{+}
+\alpha_2B_{-}}\ln{z}\ ,
\quad
k_{V}=\frac{\alpha_3 B_{+}R_{+}+\alpha_4 B_{-}R_{-}}{\alpha_3B_{+}
+\alpha_4B_{-}}\ln{z}\ ,
\label{kv}
\ee
which clearly exhibits gravito-photon mixing. Thus in general, the Love number depends on the ratio of the amplitudes on the horizon.
Following \cite{Pereniguez:2021xcj}, when computing the Love number for a specific mode, we set the tidal field of the other mode to zero, thereby fixing the relative amplitude between the two modes. Thus, when calculating $k_{\mathrm{RW}}$ ($k_{V}$), we set the coefficient of the $z^{-\tilde{\ell}}$ term in $\Psi_V|_{z\rightarrow 0}$ ($\Psi_{\mathrm{RW}}|_{z\rightarrow 0}$) to zero, fixing the ratio of the amplitudes to be $B_{+}/B_{-} = -\alpha_4/\alpha_3$ ($B_{+}/B_{-} = -\alpha_2/\alpha_1$). This allows us to derive $k_{\mathrm{RW}}$ and $k_V$ analytically.
A special case happens in $D=4$, as one finds $R_{\pm}$ vanishes. We verify that the tidal field $\Phi_{\rm tidal\pm}$ is given by polynomials of degree less than $2\tilde{\ell}+1$ and thus $k_{\mathrm{RW}}=k_V=0$.  In $D>4$, $R_\pm$ is in general nonvanishing. For $D=5$, we list values of the Love numbers (logarithm omitted) for various  $\ell$ in Fig.\ref{figkRWD5}.

\begin{figure}[ht]
    \centering
    \includegraphics[width=0.49\linewidth]{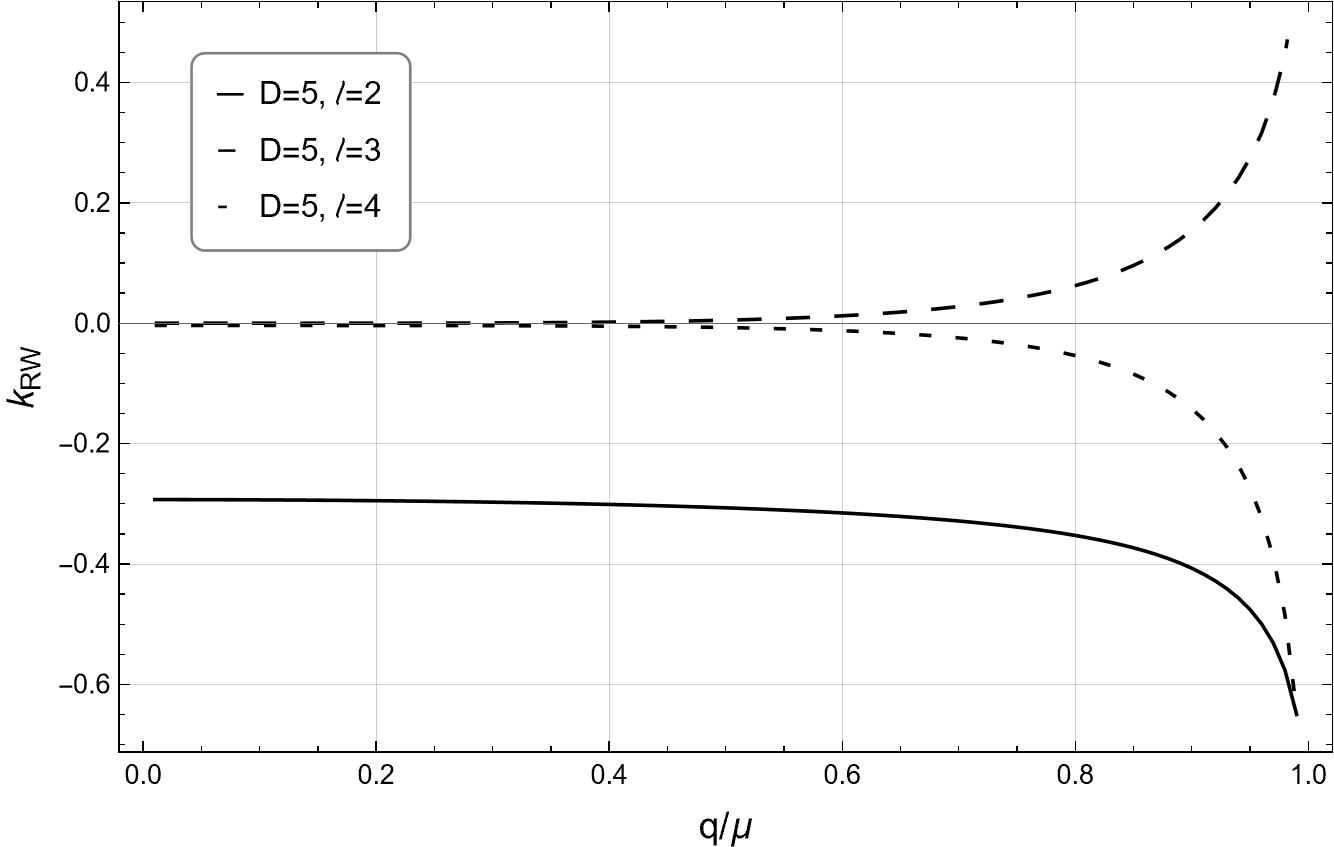}
     \includegraphics[width=0.49\linewidth]{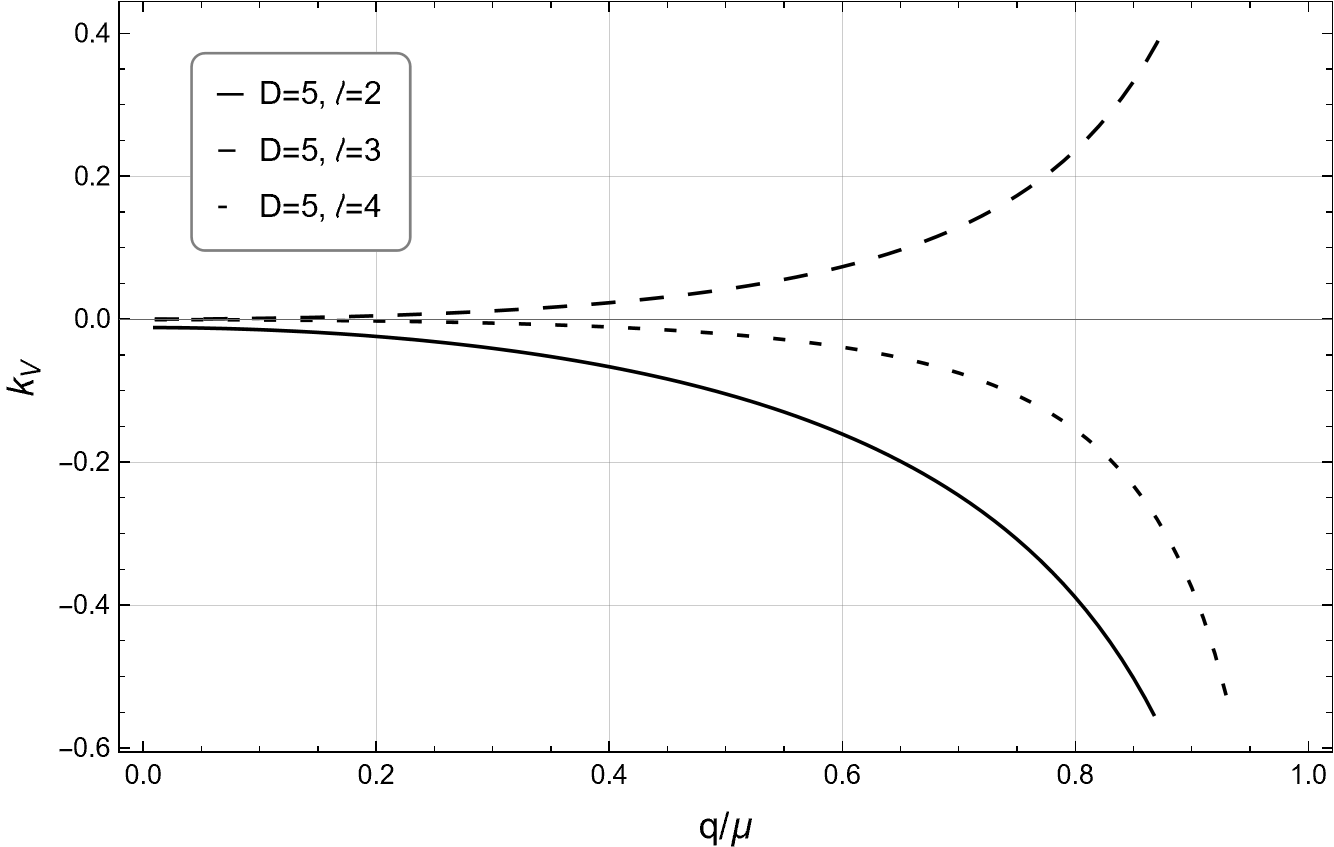}
    \caption{These plots show how vector type Love numbers in $D=5$ behave as we change $q/\mu$ and $\ell$. $\ln z$ is omitted in the plots.  }
    \label{figkRWD5}
\end{figure}

\item $2\tilde{\ell}+1 \notin \mathbb{N}$: In this case, logarithmic term is absent in \eqref{Rw}. Near the boundary at $z=0$, the general solution is given by
\bea
R_{\mathcal{W}\pm}(z) = B_{\pm} \left(k_\pm z^{\tilde{\ell}+1} \Psi_{\text{resp}\pm}(z) + z^{-\tilde{\ell}}\Psi_{\text{tidal}\pm} (z)\right)\ ,
\eea
where $\Psi_{\rm resp}$ and $\Psi_{\rm tidal}$ are analytic functions in a neighborhood of $z = 0$.
Regularity at the horizon constrains the relative normalization between the two solutions. As discussed above, the Love number for one mode is computed by setting the tidal field of the other mode to zero. Thus, we obtain
\be
k_{\mathrm{RW}}=\frac{\alpha_2 \alpha_3 k_--\alpha_1 \alpha_4 k_+}{\alpha_2 \alpha_3-\alpha_1 \alpha_4}\ ,\quad
k_{V}=\frac{ \alpha_1 \alpha_4 k_--\alpha_2 \alpha_3k_+}{\alpha_1 \alpha_4-\alpha_2 \alpha_3}\ .
\ee
For given $D$ and $\ell$, the values of $k_{\mathrm{RW}}$ and $k_{V}$ are computed numerically as before. Their dependence on $q/\mu$ is shown in Fig.\ref{figkRW} for $D=10,\ell=2, 3, 4$. Our results are consistent with those given in \cite{Pereniguez:2021xcj}. When $q=0$, our results agree with the previously known behavior for Schwarzschild black hole \cite{Hui:2020xxx}.  On the other hand, different from the neutral case, non-zero vector Love numbers are obtained for $D>5$ even when $\ell=n(D-3)\pm1$, where $n\in\mathbb{N}^+$ \cite{Pereniguez:2021xcj}. We show a few examples in this case in Fig.\ref{figkRWD68}.

\begin{figure}[ht]
    \centering
    \includegraphics[width=0.49\linewidth]{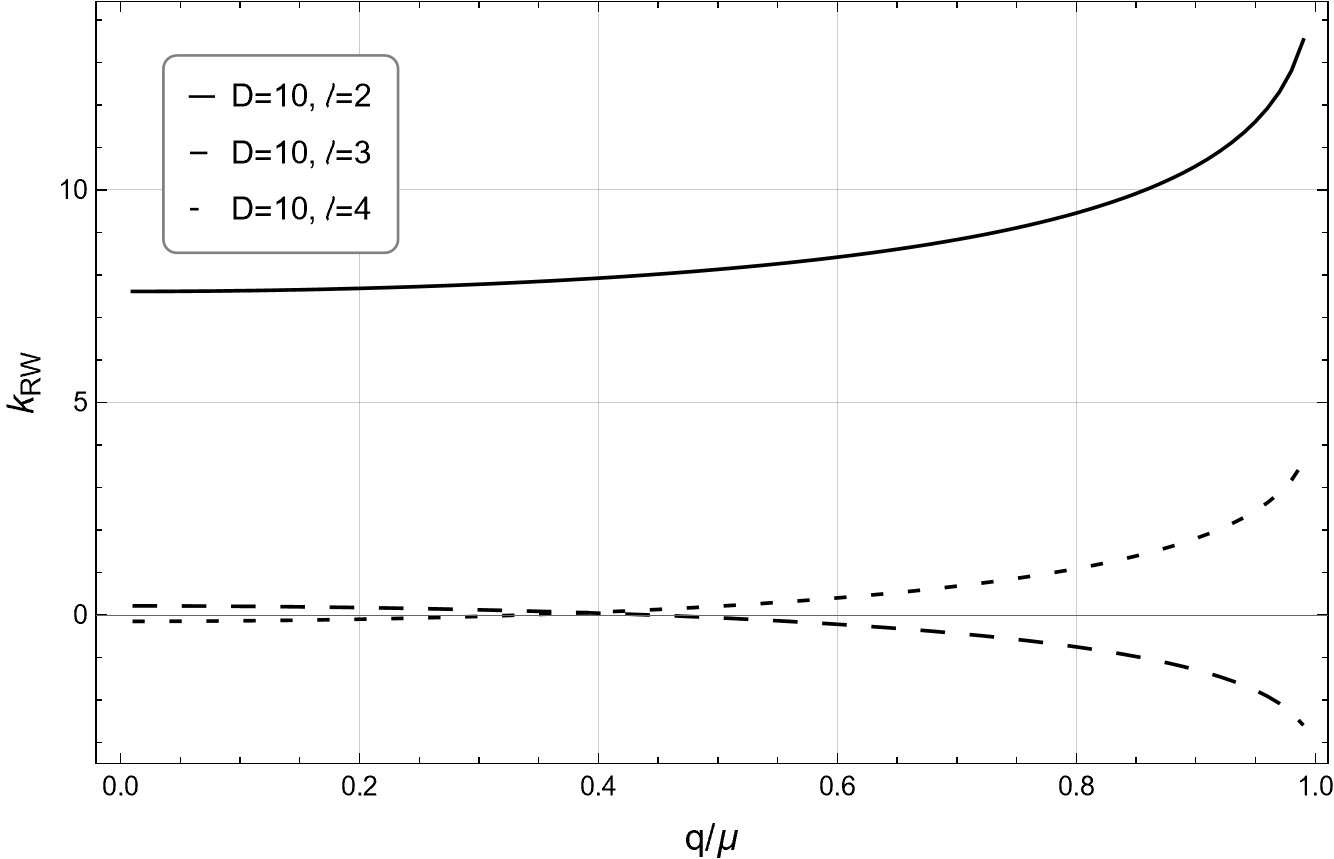}
     \includegraphics[width=0.49\linewidth]{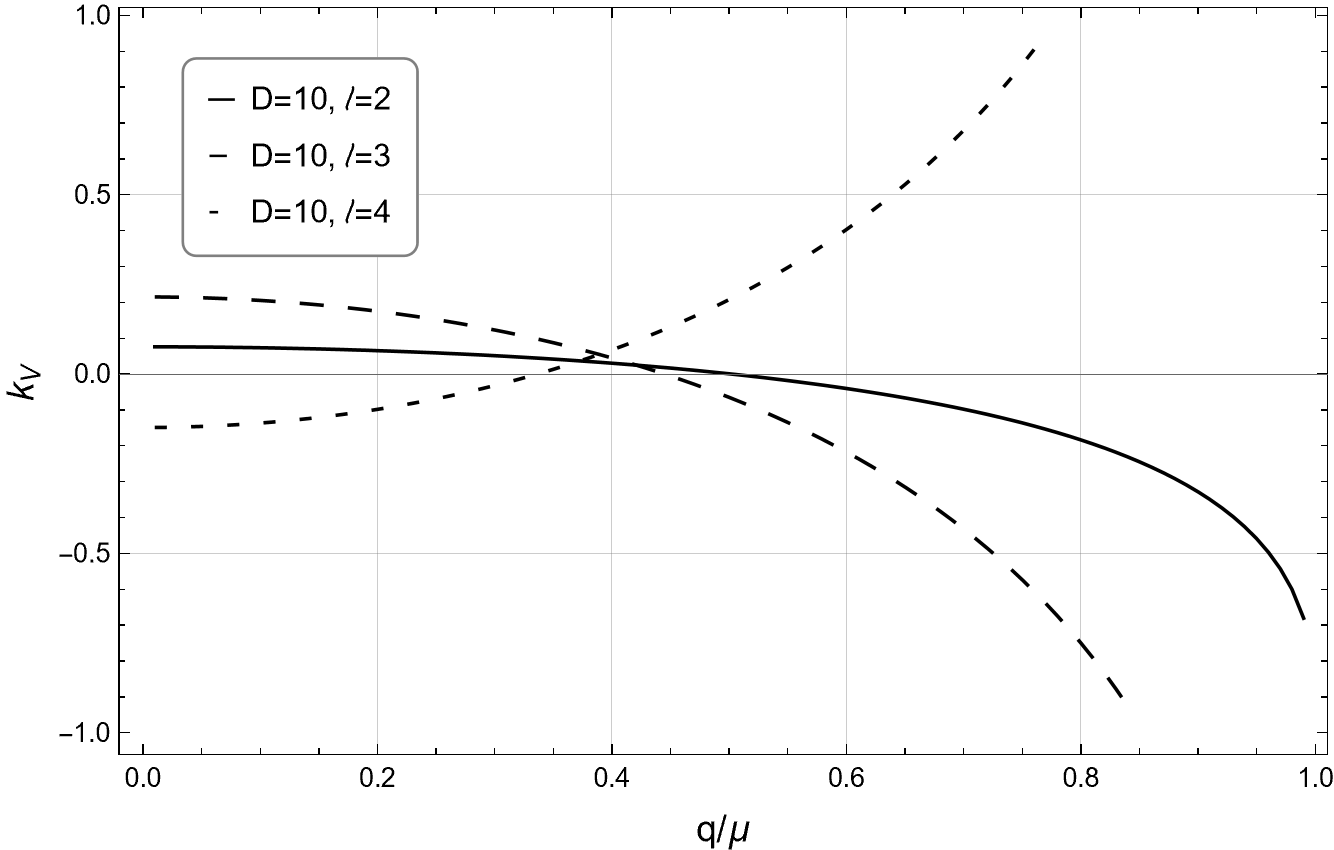}
    \caption{These plots show how vector type Love numbers in $D=10$ behave as we change $q/\mu$ and $\ell$.}
    \label{figkRW}
\end{figure}

\begin{figure}[ht]
    \centering
    \includegraphics[width=0.49\linewidth]{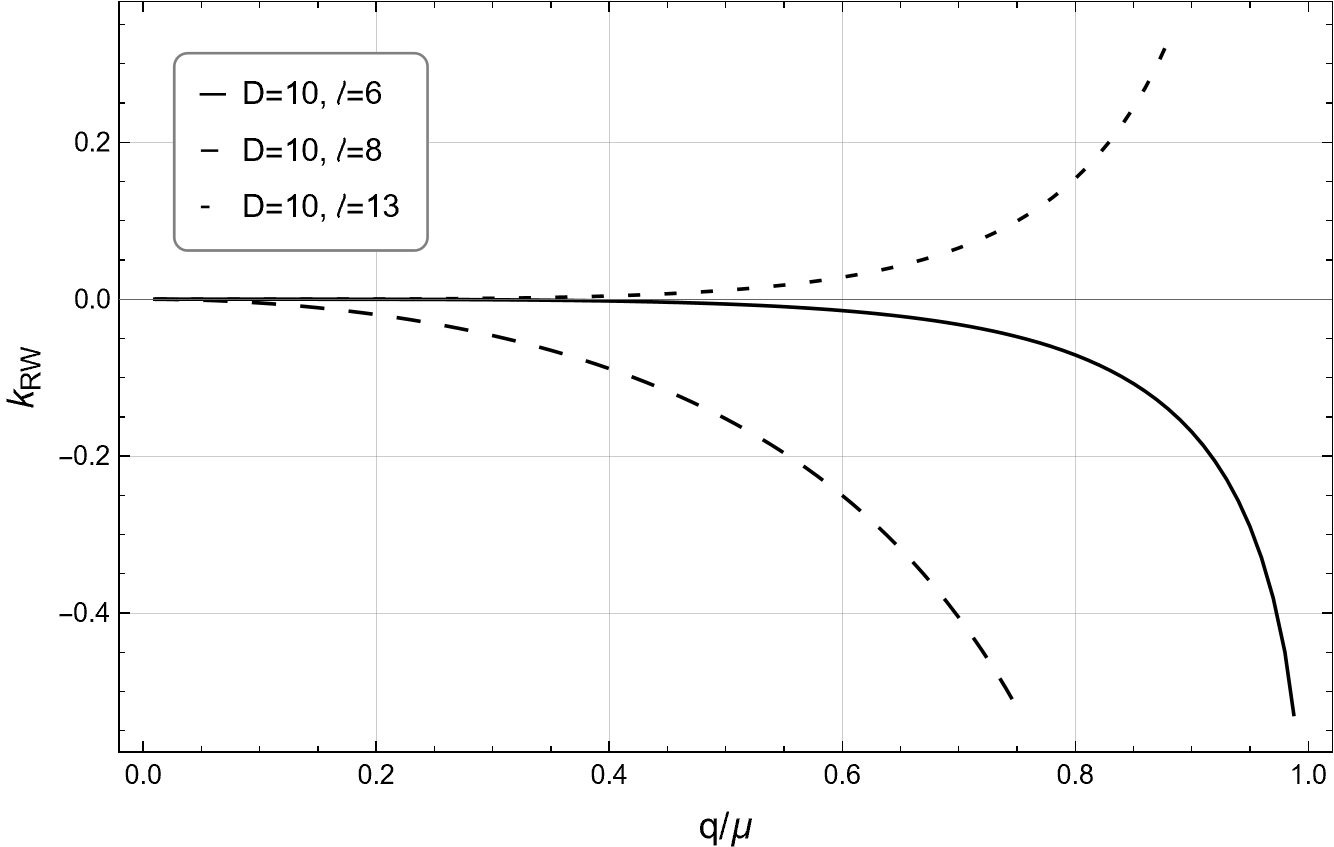}
    \includegraphics[width=0.49\linewidth]{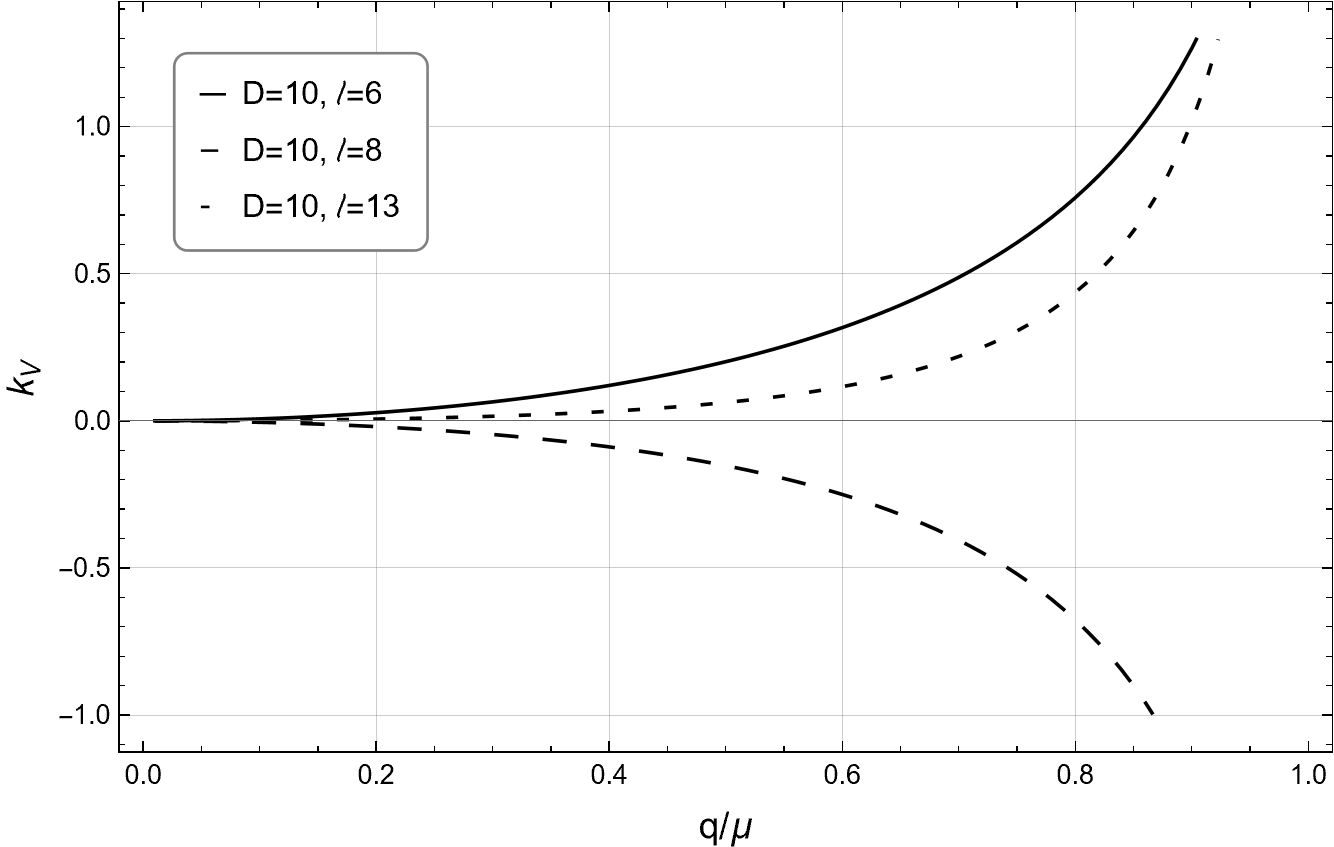}
    \caption{These plots show that vector Love numbers for RN black hole are generically non-zero in $D>5$ even when $\ell=n(D-3)\pm1$, where $n\in\mathbb{N}^+$, which is different from Schwarzschild black hole.}
    \label{figkRWD68}
\end{figure}

\end{enumerate}

\subsection{Effective action of scalar type perturbations}
Analogously to the case of Schwarzschild black hole, the metric perturbation \( h_{\mu\nu} \) and the Maxwell field perturbation \( a_{\mu} \) both contribute to the scalar-type perturbations.
From \eqref{MWaction}, \eqref{FP action} and \eqref{linear mix}, we collect the relevant terms in the effective action of scalar perturbations
\be
\mathcal{L}^{(s=0)}=\widetilde{\mathcal{L}}_h^{(s=0)}+\widetilde{\mathcal{L}}_a^{(s=0)}+\widetilde{\mathcal{L}}_m^{(s=0)}\ .
\label{scalar sector}
\ee
In order to eliminate the redundant degrees of freedom, we introduce the auxiliary field $P_{\rm aux}$ and replace part of the action in as follows
\be
\Big(\dot{a}_r-a_t'+\frac{1}{2}\psi'(H_2-H_0)\Big)^2\longrightarrow
\Big(2P_\mathrm{aux}\big(\dot{a}_r-a_t'+\frac{1}{2}\psi'(H_2-H_0)\big)
-P_\mathrm{aux}^2\Big)\ .
\label{Ag s0}
\ee
Next, rather than solving for $P_{\rm aux}$, we solve for \(a_t\) and \(a_r\) using their equations of motion  in terms of \(P_{\mathrm{aux}}\)
\bea
a_t&=&f\mathcal{H}_1\psi'-\frac{rf}{\ell(\ell+D-3)}\big[(D-2)P_{\mathrm{aux}}+rP_{\mathrm{aux}}'\big],
\nn\\
a_r&=&-\frac{r^2}{\ell(\ell+D-3)f}\dot{P}_{\mathrm{aux}}\ ,
\eea
and substitute the expressions above back to \eqref{scalar sector}. Since
$H_1$ obeys an algebraic equation, we can solve it easily
\bea
H_1=\dot{\mathcal{H}}_1-\frac{r(D-2)}{\ell(\ell+D-3)}\dot{H}_2\ ,
\eea
which will again be plugged in back to the action.
We further trade  $\mathcal{H}_1$ for another variable
$\mathcal{V}$
\be
\mathcal{V}=\mathcal{H}_1-\frac{r(D-2)}{2\ell(\ell+D-3)}H_2\ ,
\ee
so that we can solve $H_2$ in terms of $\mathcal{V}$ from $H_0$ equation of motion
\bea
H_2&=&-\frac{2 \ell  (\ell+D -3)}{2 \ell  (\ell+D -3)-2 (D-2) f+(D-2) r f'}
\nn\\
&&\times\Big(
\frac{r^2\psi'}{\ell  (\ell+D -3)}P_{\rm{aux}}-f'\mathcal{V}-\frac{2f}{r}\big((D-3)\mathcal{V}+r\mathcal{V}'\big)
\Big)\ .
\eea
Eventually, the scalar sector action reduces to one involving only two variables, \( \mathcal{V} \) and \( P_{\mathrm{aux}} \), which are coupled together. To transform the action into standard form, we define \( \{\Psi_\mathrm{Z}, \Psi_S\} \)
\bea
\Psi_S&=&\frac{r^{\frac{D}{2}}}{\sqrt{\ell(\ell+D-3)}}P_{\rm{aux}}\ ,
\nn\\
\Psi_{\mathrm{Z}}&=&\frac{2  \sqrt{\ell(D-2)(\ell+D-3){\cal F}} r^{\frac{D-4}{2}}  f }{\mathcal{G}}\mathcal{V}\ ,
\eea
where the function $\cal F$ and $\cal G$ are
\bea
\mathcal{F}&=&-2 (D-2) r f'-2 (D-3) (D-2) f+(D-3) \big(2 \ell  (\ell+D -3)+r^2 \psi '^2\big)\ ,
\nn\\
\mathcal{G}&=&(D-2) r f'-2 (D-2) f+2 \ell  (\ell+D -3)
\eea
Now the scalar sector action becomes
\bea
\mathcal{L}^{(s=0)}&=&\mathcal{L}_{\mathrm{Z}}+\mathcal{L}_S+\mathcal{L}_{\mathrm{mix}}\ ,
\nn\\
\mathcal{L}_{\mathrm{Z}}&=&\frac{1}{2f}\dot{\Psi}_{\mathrm{Z}}^2-\frac{f}{2}\Psi_{\mathrm{Z}}'^2-\frac{1}{2}V_{\mathrm{Z}}(r)\Psi_{\mathrm{Z}}^2\
\nn\\
\mathcal{L}_S&=&\frac{1}{2f}\dot{\Psi}_S^2-\frac{f}{2}\Psi_S'^2-\frac{1}{2}V_{S}(r)\Psi_S^2\ ,
\nn\\
\mathcal{L}_{m}&=&-\frac{2\sqrt{D-2}r\psi'}{\sqrt{\mathcal{F}}}\Big(
\frac{1}{2f}\dot{\Psi}_S\dot{\Psi}_{\mathrm{Z}}-\frac{f}{2}\Psi_S'\Psi_{\mathrm{Z}}'
-\frac{1}{2}V_{m1}(r)\Psi_S\Psi_{\mathrm{Z}}'-\frac{1}{2}V_{m2}(r)\Psi_S\Psi_{\mathrm{Z}}
\Big).\label{action S Z}
\eea
The explicit forms of $V_\mathrm{Z}, V_S, V_{m1}, V_{m2}$ are complicated and we postpone them to the appendix \eqref{pentential ZS}. In the vector sector \eqref{action RW V}, the two modes entangle with each other via a mixed mass term. In contrast, here the mixing terms contain derivatives of fields, making the decoupling process more complicated.

The gauge-invariant scalar type Love numbers are determined from the $C_{trtr}$ component of Weyl tensor and  the temporal component of  the Maxwell field  $A_t$. We find that they are related to $\Psi_{\mathrm{Z}}$ and $\Psi_S$ via \cite{Hui:2020xxx}
\be
A_t=\frac{D+2\ell-4}{2\sqrt{\ell(\ell+D-3)}}r^{\frac{2-D}{2}}\Psi_SY_{\ell m},\quad C_{trtr}\xrightarrow{r\rightarrow\infty}2(D-3)\ell(\ell+D-3)r^{\frac{D}{2}-5}\Psi_{\mathrm{Z}} Y_{\ell m}\ .
\ee
The Love numbers obtained from \(\Psi_S\) and \(\Psi_{\mathrm{Z}}\) are commonly referred to as the electric response coefficient and the gravito-electric response coefficient.

We now move on to diagonalize the scalar type perturbations. We make the ansatz
\be
\Psi_S=a_1(r)\mathcal{U}_{+}+a_2(r)\mathcal{U}_{-},\quad \Psi_{\mathrm{Z}}=a_3(r)\mathcal{U}_{+}+a_4(r)\mathcal{U}_{-}\ ,
\label{decouple scalar0}
\ee
and try to find $a_1,\cdots a_4$ so that there is no mixing between ${\cal U}_{\pm}$.
Different from the vector sector \eqref{decouple vector},  here $a_1,\cdots a_4$ are functions of \( r \) rather than constants. Their explicit forms are
\bea
a_1&=&-\frac{c_2 (D-2) q +(\ell -1)  (\ell+D -2)r^{D-3}}{r^{D-3}\sqrt{(\ell -1)(\ell+D -2)} \sqrt{c_2^2+(\ell -1) (\ell+D -2)}}\ ,
\nn\\
a_3&=&\frac{c_2 \sqrt{\mathcal{F}}}{\sqrt{2(D-3)} \sqrt{(\ell -1)(\ell+D -2)} \sqrt{c_2^2+(\ell -1) (\ell+D -2)}}\ ,
\nn\\
a_2&=&\frac{c_2 r^{D-3}+(2-D) q }{r^{D-3}\sqrt{c_2^2+(\ell -1) (\ell+D -2)}},\quad a_4=\frac{\sqrt{\mathcal{F}}}{\sqrt{2(D-3)} \sqrt{c_2^2+(\ell -1) (\ell+D -2)}}\ ,
\label{decouple scalar}
\eea
where the constant $c_2$ is
\be
c_2=-\frac{2  (\ell -1) (\ell+D -2)q}{(D-1) \mu +\sqrt{(D-1)^2 \mu ^2+4  (\ell -1) (\ell+D -2)q^2}}\ .
\label{c2}
\ee
and then we decouple the action \eqref{action S Z} successfully
\bea
\mathcal{L}^{(s=0)}&=&\mathcal{L}_{\mathcal{U}+}+\mathcal{L}_{\mathcal{U}-}\ ,
\nn\\
\mathcal{L}_{\mathcal{U}+}&=&\frac{1}{2f}\dot{\mathcal{U}}_{+}^2-\frac{f}{2}\mathcal{U}_{+}'^2-\frac{1}{2}V_{\mathcal{U}+}(r)\mathcal{U}_{+}^2\ ,
\nn\\
\mathcal{L}_{\mathcal{U}-}&=&\frac{1}{2f}\dot{\mathcal{U}}_{-}^2-\frac{f}{2}\mathcal{U}_{-}'^2-\frac{1}{2}V_{\mathcal{U}-}(r)\mathcal{U}_{-}^2\ .
\label{scalar sector action decouple}
\eea
where the potentials in equations above are postponed to the appendix \eqref{pentential U+} and \eqref{pentential U-} . For \(D = 4\), we find that the equations of motion for \(\mathcal{U}_{\pm}\) agree with those  in \cite{Moncrief:1974ng} where the corresponding functions are denoted by $R_\pm$. For Schwarzschild black hole ($q=0$) the coefficients in \eqref{decouple scalar} reduce to
\be
a_1=-1,\quad a_2=a_3=0,\quad a_4=1\ .
\label{decouple scalar1}
\ee

\subsection{Scalar-type Love numbers in diverse dimensions}
We now proceed to solve the decoupled scalar modes \eqref{scalar sector action decouple}.
After applying separation of variables
\be
\mathcal{U}_{\pm}(t,r)=e^{-i\omega t}r^{\frac{D-2}{2}}R_{\mathcal{U}\pm}(r)\ ,
\ee
and taking the static limit $\omega \rightarrow 0$, the equations for the radial functions become
\bea
R_{\mathcal{U}\pm}'' + \frac{\left((D-2) f + r f'\right)}{r f} R_{\mathcal{U}\pm}' + \frac{ (D^2-6D+8) f +2 (D-2) r f' - 4 r^2 V_{\mathcal{U}\pm}}{4 r^2 f} R_{\mathcal{U}\pm} = 0\ .
\eea
Again, in terms of the variable $z =\left(\frac{r_+}{r} \right)^{D-3}$, near spatial infinity at $z=0$, the solutions are characterized by exponents $-\tilde{\ell}$ and $\tilde{\ell} + 1$. Similar to the vector case, we consider two subcases:
\begin{enumerate}
    \item  $2\tilde{\ell}+1 \in \mathbb{N}$: The general solution in this case is
\bea
R_{\mathcal{U}\pm}(z) = B_{\pm}\left(k_\pm z^{\tilde{\ell}+1} \Psi_{\text{resp}\pm}+z^{-\tilde{\ell}}\Psi_{\text{tidal}\pm} + R_{\pm} z^{\tilde{\ell}+1} \Psi_{\text{resp}\pm} \ln z\right).
\eea
The unambiguous Love number is usually defined using coefficients $R_\pm$ (when it is non-vanishing) which can be computed analytically by substituting the expansion into the equations. Accordingly, the asymptotic behaviors of $\Psi_{S}$ and $\Psi_{\mathrm{Z}}$ near the boundary are given by
\bea
\Psi_{S}|_{z\rightarrow 0} &= (a_1 B_+ + a_2 B_-) z^{-\tilde{\ell}} + (a_1 B_+ R_+ + a_2 B_- R_-) z^{\tilde{\ell}+1} \ln z\ ,
\\
\Psi_{\mathrm{Z}}|_{z\rightarrow 0} &= (a_3 B_+ + a_4 B_-) z^{-\tilde{\ell}} + (a_3 B_+ R_+ + a_4 B_- R_-) z^{\tilde{\ell}+1} \ln z\ .
\eea
We then define the Love number for one mode by turning off the tidal field of the other mode and obtain
\bea
&&k_S=\frac{a_1\lambda_SR_++a_2R_- }{a_1\lambda_S+a_2}\ln z=\frac{c_2^2 R_-+(\ell-1) (D+\ell-2) R_+}{c_2^2+(\ell-1) (D+\ell-2)}\ln z
\ ,
\\
&&k_{\mathrm{Z}}=\frac{a_3\lambda_ZR_++a_4R_-}{a_3\lambda_Z+a_4}\ln z=\frac{c_2^2 R_++(\ell-1) (D+\ell-2) R_-}{c_2^2+(\ell-1) (D+\ell-2)}\ln z
\ .
\eea
Here, $\lambda_S = -a_4/a_3$ and $\lambda_{\mathrm{Z}} = -a_2/a_1$ are the relative amplitudes determined by setting the one of the source term to zero, and $c_2$ is defined in \eqref{c2}.

We find that for $\tilde{\ell}\in \mathbb{N}$, $R_\pm$ vanishes. However, unlike the tensor and vector cases, the scalar tidal field $\Psi_{\rm tidal_\pm}$ becomes an infinite polynomial rather than a polynomial of degree $<2\tilde{\ell}+1$. To isolate the tidal and possible response terms, we shift $\tilde{\ell}$ away from integer by a small amount $\epsilon$.  Via numerical calculations, we find that as $\epsilon$ approaches zero, $k_\pm$ in front of the response term also tends to zero. We exhibit the results  in Fig.\ref{figkSD4epsilon} for $D = 4,5,6$. In particular, we confirm that in $D = 4$, the solution reduces to pure tidal field term and the scalar type Love numbers vanish \cite{Rai:2024lho}.
\begin{figure}[ht!]
    \centering
    \includegraphics[width=0.49\linewidth]{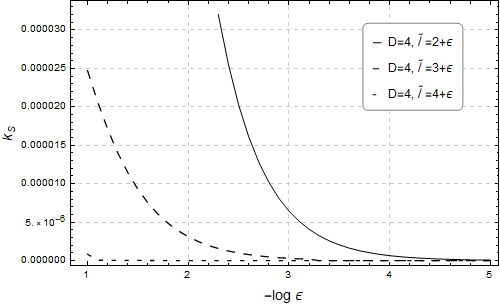}
    \includegraphics[width=0.49\linewidth]{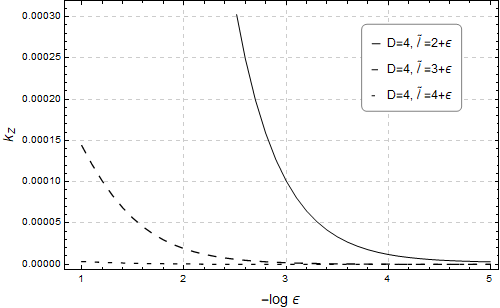}
     \includegraphics[width=0.49\linewidth]{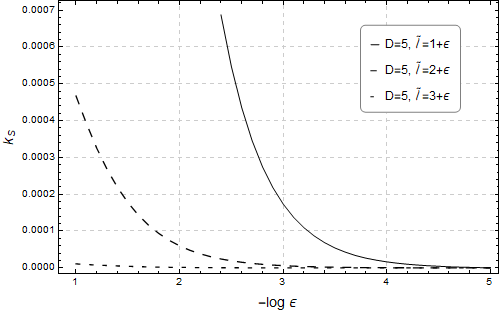}
    \includegraphics[width=0.49\linewidth]{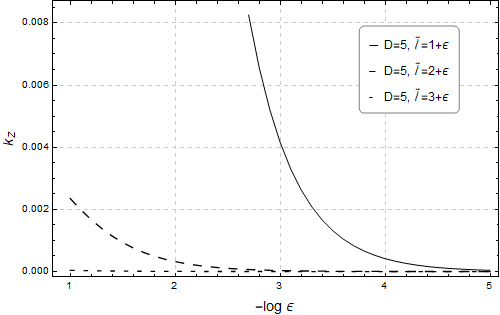}
    \includegraphics[width=0.49\linewidth]{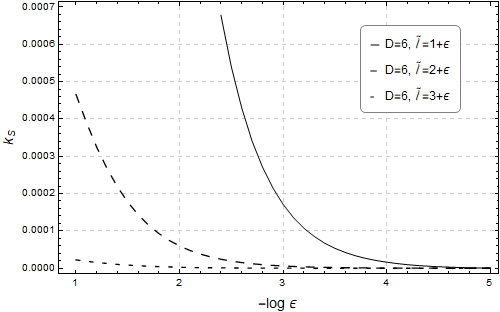}
    \includegraphics[width=0.49\linewidth]{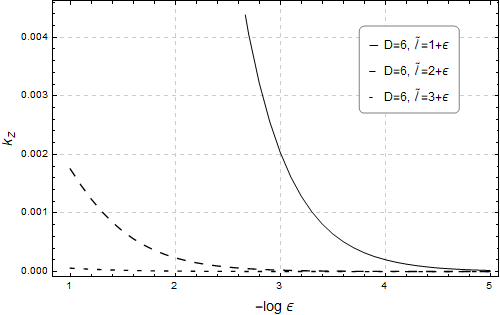}
    \caption{In these plots, we show that when $\tilde{\ell}\in \mathbb{N}, q/\mu=0.1$, the scalar type Love numbers of RN black hole vanish. Examples above correspond to $D=4,5,6$ and $\epsilon$ ranges from $10^{-1}$ to $10^{-5}$.}
    \label{figkSD4epsilon}
\end{figure}
For $\tilde{\ell}\in\frac{1}{2}+\mathbb{N}$, $R_\pm$ takes non-zero real value and the Love numbers exhibit logarithmic running behaviors. We show the results in Fig. \ref{figkSD5} for the the case of $D=5$ and $\ell=3,5,7$.
\begin{figure}[ht!]
    \centering
    \includegraphics[width=0.49\linewidth]{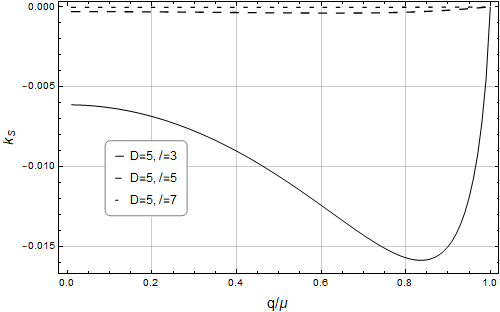}
    \includegraphics[width=0.49\linewidth]{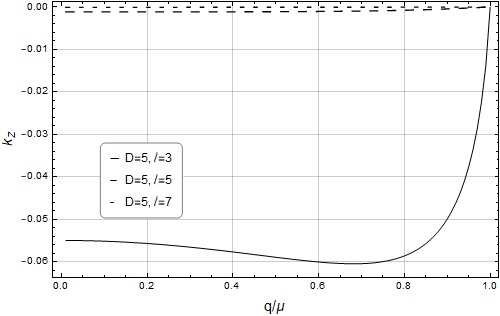}
    \caption{Scalar Love numbers for RN black hole in $D=5$ with $\ell=3,5,7$, where $\ln z$ is omitted.}
    \label{figkSD5}
\end{figure}
When switching off the electric charge, our results agree with known results for Schwarzschild black hole \cite{Hui:2020xxx}.

\item $2\tilde{\ell}+1 \notin \mathbb{N}$:  In this case, the two branches are linearly independent, therefore $R_\pm=0$. The
general solution is of the form
\be
R_{\mathcal{U}\pm}(z) = B_{\pm}\left(k_\pm z^{\tilde{\ell}+1} \Psi_{\text{resp}\pm}+z^{-\tilde{\ell}}\Psi_{\text{tidal}\pm} \right).
\ee
The love numbers are defined as
\be
k_S=\frac{c_2^2 k_-+(\ell-1) (D+\ell-2) k_+}{c_2^2+(\ell-1) (D+\ell-2)},\quad
k_{\mathrm{Z}}=\frac{c_2^2 k_++(\ell-1) (D+\ell-2) k_-}{c_2^2+(\ell-1) (D+\ell-2)} \ .
\ee
The values of $k_{S}$ and $k_{\mathrm{Z}}$ are computed via numerical method as before for given $D$ and $\ell$. Examples corresponding to $D=10$ and $\ell=2, 3, 4$ are shown in Fig.\ref{figkS}. In the neutral limit $q\rightarrow0$, the numerical results agree with the one in \cite{Hui:2020xxx}.
\begin{figure}[ht]
    \centering
    \includegraphics[width=0.49\linewidth]{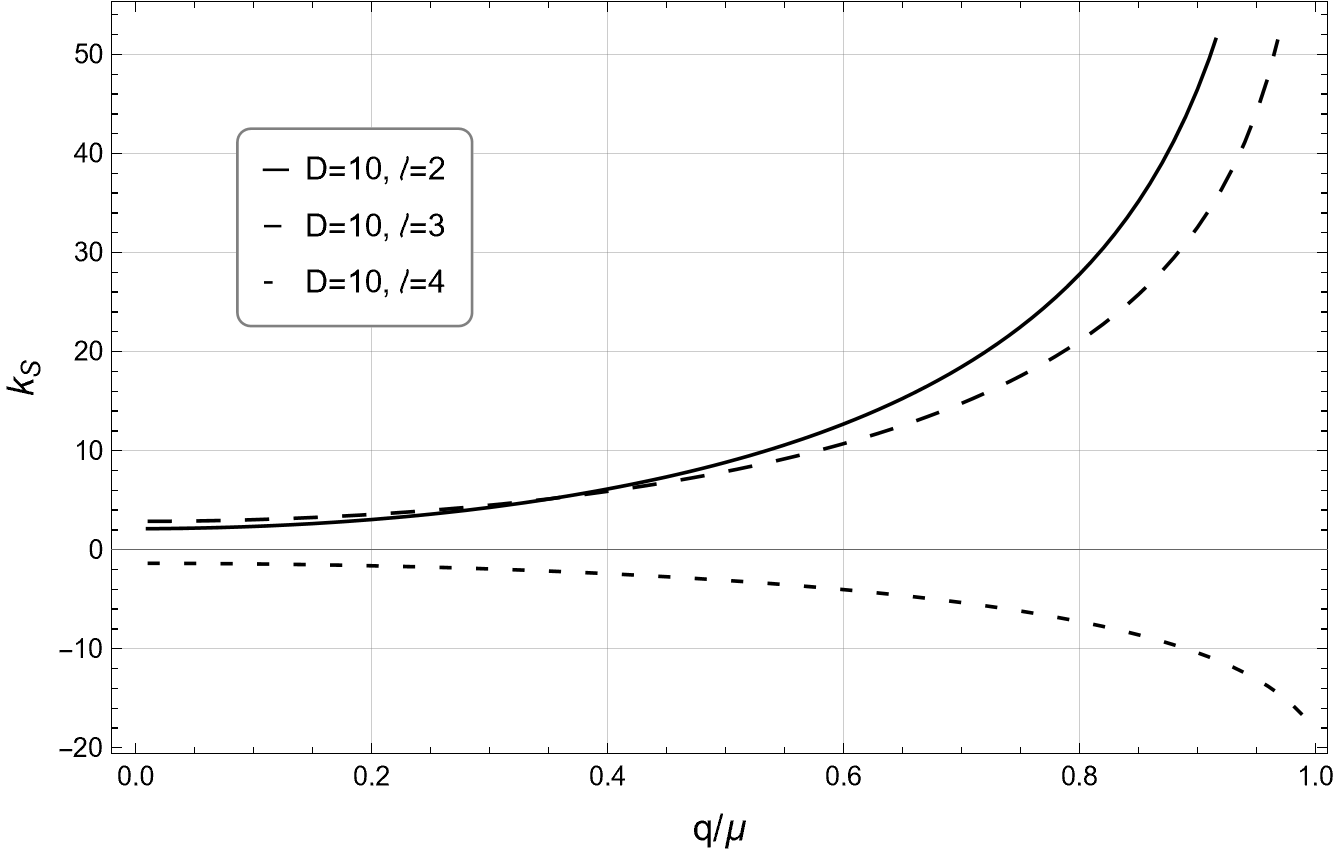}
     \includegraphics[width=0.49\linewidth]{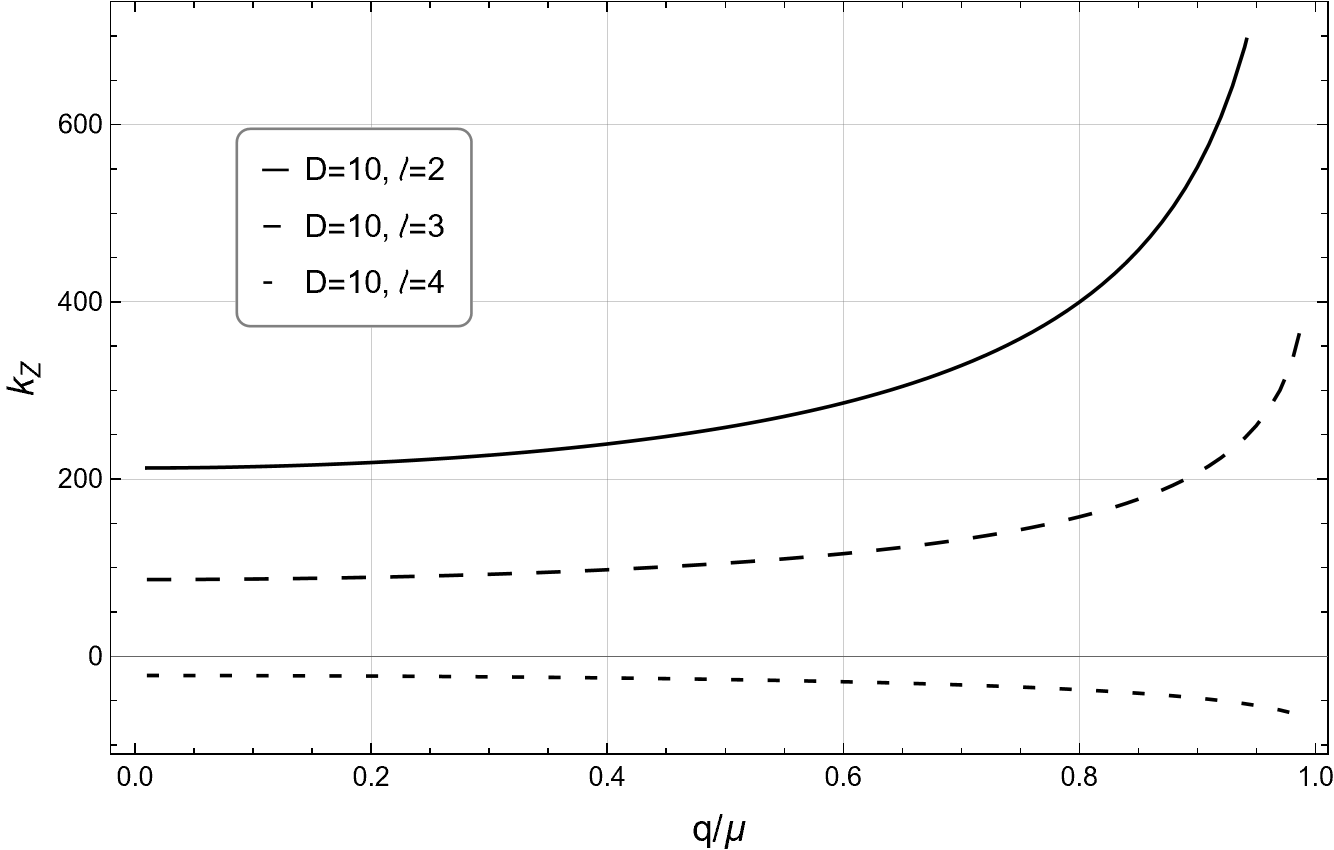}
    \caption{Scalar Love numbers for $D=10$ RN black holes with $\ell=2,3,4$.}
    \label{figkS}
\end{figure}

\end{enumerate}

\section{Conclusions}

In this paper, we computed the tidal Love numbers of $D$ dimensional RN black holes in all perturbation channels. We obtained the effective action for the perturbations and successfully diagonalized them to arrive at a set of master equations. We showed that tensor perturbations are governed by hypergeometric functions and the vector perturbations satisfy Heun type equations in general. The scalar perturbations obey more complicated equations which might be reduced to simpler equations using the trick employed in $D=4$, which deserves further study in the future.

From the master equations, we extracted Love numbers in various sector. In the tensor and vector sectors, we recovered known results in the literature. In the scalar sector, we observed that the Love numbers vanish when the effective multipolar index $\tilde{\ell}\in\mathbb{N}$, analogous to the Schwarzschild black hole. It is possible that the vanishing of tidal Love numbers in higher dimensions can be associated with certain accidental symmetry.  As expected, when
$\tilde{\ell}\in\mathbb{N}+\frac12$, the scalar Love numbers exhibit logarithmic running behaviors and when $\tilde{\ell}\notin\mathbb{N}$, the scalar Love numbers are finite real values.

In this work, we focused on the purely electric RN black hole. It is conceivable that via electromagnetic duality, our results can be generalized to dyonically charged black hole. We would like to consider spherically symmetric black hole in Einstein-Maxwell-dilaton system, some of which do admit analytically solutions. It should also be interesting to consider nonlinear tidal Love numbers for charged black holes as was done for the neutral ones \cite{DeLuca:2023mio}. In \cite{Cvetic:2021vxa}, the Love numbers of four-dimensional rotating STU black
holes were analyzed. These results show that, for black holes in string theory and supergravity,
the behavior of the Love numbers at the two-derivative level matches that of conventional black
holes (e.g., Schwarzschild black holes). However, when higher-derivative terms are included,
the stringy parameter $\alpha'$ will also contributes to the Love numbers and potentially lift up the degeneracy. Thus we would also like to investigate Love numbers of black holes in string theory with $\alpha'$ corrections \cite{Ma:2025vnr,Hu:2025aji, Ma:2021opb, Ma:2022nwq, Ozkan:2024euj}. Finally, our techinques may be generalized to study linear response of charged AdS black branes \cite{Ge:2008ak} or holes, which may lead to interesting applications in holography in diverse dimensions.

\section*{Acknowledgement}
L.~Ma and Y.~Pang are supported by the National Key R\&D Program No. 2022YFE0134300 and the National Natural Science Foundation of China (NSFC) under Grant No. 12575076, No.12247103.
L.~Ma is also supported in part by NSFC grant No.~12447138, Postdoctoral Fellowship Program of CPSF Grant No.~GZC20241211 and the China Postdoctoral Science Foundation under Grant No.~2024M762338. H.~L\"u is supported in part by NSFC grants No.~12375052 and No.~11935009. The work is also supported by the Tianjin University Self-Innovation Fund Extreme Basic Research Project Grant No.~2025XJ21-0007.

\paragraph*{Conflict of Interest\quad The authors declare that they have no conflict of interest.}

\section*{Appendix  Explicit forms of the potentials in the scalar sector}

Here, we present the potential in \eqref{action S Z}
\bea
V_S&=&\frac{D\big((D-2) f+2 r f'\big)}{4 r^2}+\frac{1}{\mathcal{G}^2 r^2}\Bigg[
-4 (D-2)^3 f^3-(D-2)^3 r^3 f'^3\cr
&&+4 \ell ^2 (\ell+D -3)^2 \big(\ell  (\ell+D -3)+r^2 \psi '^2\big)-(D-2)^2 r^2 f'^2 \big(3 \ell  (\ell+D -3)+r^2 \psi '^2\big)\cr
&&+4 (D-2)^2 f^2 \big(3 \ell  (\ell+D -3)-2 (D-3) r^2 \psi '^2\big)
+(D-2) f \Big(2r^2 \psi '^2 \big((D-4) (D-2) r f'\cr
&&+2 (2 D-7) \ell  (\ell+D -3)\big)+3 (D-2)^2r^2 f'^2+2 (D-3) r^4 \psi '^4-12 \ell ^2 (\ell+D -3)^2\Big)
\Bigg]\ ,
\nn\\
V_{\mathrm{Z}}&=&\frac{1}{4 (D-2) \mathcal{G}^2 r^2 \mathcal{F}^2}\Bigg[8 (D-2) f \mathcal{F}^4-8 (D-2) \mathcal{G}^4 \big(2 (D-2)^2 f+\mathcal{F}\big)
+8 \mathcal{G} \mathcal{F}^2 \Big(\cr
&&4 (D-1) (D-2)^3 f^2+(D-2)^2 f \big(\mathcal{F}-4 (D-1) \ell  (D+\ell -3)\big)+2 \mathcal{F} \ell  (D+\ell -3)\Big)
\cr
&&+2 (D-2) \mathcal{G}^3 \Big(8 (D-2) f \big(2 (D-2) (D-1) \ell  (D+\ell -3)-D \mathcal{F}\big)-16 (D-1) (D-2)^3 f^2\cr
&&+\mathcal{F} \big(4 (D+3) \ell  (D+\ell -3)-\mathcal{F}\big)\Big)+\mathcal{G}^2 \Big(8 (D-2)^3 f^2 \big(4 (D-2) (D-1)^2 \ell  (D+\ell -3)
\cr
&&-(D^2-1) \mathcal{F}\big)-16 (D-1)^2 (D-2)^5 f^3-(D-2)^2 f \big(-8 (D-1) (D+5) \mathcal{F} \ell  (D+\ell -3)\cr
&&-(D+20) \mathcal{F}^2+16 (D-2) (D-1)^2 \ell ^2 (D+\ell -3)^2\big)+4 \mathcal{F} \big(2 (D-1) \mathcal{F} \ell  (D+\ell -3)\cr
&&-8 (D-2) (D-1) \ell ^2 (D+\ell -3)^2-\mathcal{F}^2\big)\Big)
\Bigg]\ ,
\nn\\
V_{m1}&=&-\frac{(D-3) f}{r \mathcal{F}}\Big(
2 (D-2) r f'+2 (D-3) \big((D-2) f-\ell  (D+\ell -3)\big)+r^2 \psi '^2
\Big)\ ,
\nn\\
V_{m2}&=&\frac{1}{4 (D-2) \mathcal{G}^2 r^2 \mathcal{F}^2}\Bigg[8 (D-2) f \mathcal{F}^4-8 (D-2) \mathcal{G}^4 \big(6 (D-2)^2 f+\mathcal{F}\big)\cr
&&+8 \mathcal{G} \mathcal{F}^2 \Big(2 (D-1) (D-2)^3 f^2+(D-2) f \big((D-3) \mathcal{F}-2 (D-2) (D-1) \ell  (D+\ell -3)\big)\cr
&&+2 \mathcal{F} \ell  (D+\ell -3)\Big)-2 (D-2) \mathcal{G}^3 \Big(48 (D-1) (D-2)^3 f^2-\mathcal{F} \big(4 (D+1) \ell  (D+\ell -3)
\cr
&&-\mathcal{F}\big)-12 (D-2) f \big(4 (D-2) (D-1) \ell  (D+\ell -3)-D \mathcal{F}\big)\Big)
+\mathcal{G}^2 \Big((D-2)^2 f \big(D \mathcal{F}^2
\cr
&&+8 (D-1) (2 D+3) \mathcal{F} \ell  (D+\ell -3)-48 (D-2) (D-1)^2 \ell ^2 (D+\ell -3)^2\big)+8 (D-1) (D\cr
&&-2)^3 f^2 \big(12 (D-2) (D-1) \ell  (D+\ell -3)-(2 D+1) \mathcal{F}\big)-48 (D-1)^2 (D-2)^5 f^3\cr
&&+4 \mathcal{F} \big(2 (D-2) \mathcal{F} \ell  (D+\ell -3)-4 (D-2) (D-1) \ell ^2 (D+\ell -3)^2-\mathcal{F}^2\big)\Big)
\Bigg].\label{pentential ZS}
\eea

After diagonalizing the perturbations in scalar sector, the potential terms in the decoupled action \eqref{scalar sector action decouple} are given by
\bea
V_\mathcal{U+}&=&\frac{\sqrt{D-2}  \left(2 r a_3 \mathcal{F} a_1' \psi '+a_1 \left(a_3 \left(2 \mathcal{F} \left(r \psi ''+\psi '\right)-r \psi ' \mathcal{F}'\right)-2 r \mathcal{F} a_3' \psi '\right)\right)}{2 \mathcal{F}^{3/2}}V_{m1}\nonumber\\
&&+\frac{\sqrt{D-2} r a_1 a_3 \psi '}{\sqrt{\mathcal{F}}}V_{m1}'-\frac{2 \sqrt{D-2} r a_1 a_3\psi '}{\sqrt{\mathcal{F}}} V_{m2} +a_1^2 V_S+a_3^2 V_Z\nonumber\\
&&+\frac{1}{2 \mathcal{F}^{3/2}}\Bigg[a_3 \Big(2 \sqrt{D-2} \mathcal{F} \left(r f a_1' \psi ''+\psi ' \left(r f a_1''+a_1' \left(r f'+f\right)\right)\right)
\nonumber\\
&&-\sqrt{D-2} r f a_1' \psi ' \mathcal{F}'-2 \mathcal{F}^{3/2} \left(f a_3''+a_3' f'\right)\Big)+a_1 \Big(-2 \mathcal{F}^{3/2} \left(f a_1''+a_1' f'\right)\nonumber\\
&&-\sqrt{D-2}r f a_3' \psi ' \mathcal{F}'+2 \sqrt{D-2} \mathcal{F} \left(r f a_3' \psi ''+\psi ' \left(r f a_3''+a_3' \left(r f'+f\right)\right)\right)\Big)\Bigg]\ ,
\label{pentential U+}
\nn\\
V_\mathcal{U-}&=&\frac{\sqrt{D-2}  \left(2 r a_4 \mathcal{F} a_2' \psi '+a_2 \left(a_4 \left(2 \mathcal{F} \left(r \psi ''+\psi '\right)-r \psi ' \mathcal{F}'\right)-2 r \mathcal{F} a_4' \psi '\right)\right)}{2 \mathcal{F}^{3/2}}V_{m1}\nonumber\\
&&+\frac{\sqrt{D-2} r a_2 a_4 \psi '}{\sqrt{\mathcal{F}}}V_{m1}'-\frac{2 \sqrt{D-2} r a_2 a_4\psi '}{\sqrt{\mathcal{F}}} V_{m2} +a_2^2 V_S+a_4^2 V_Z
\cr
&&+\frac{1}{2 \mathcal{F}^{3/2}}\Bigg[a_4 \Big(2 \sqrt{D-2} \mathcal{F} \left(r f a_2' \psi ''+\psi ' \left(r f a_2''+a_2' \left(r f'+f\right)\right)\right)
\nonumber\\
&&-\sqrt{D-2} r f a_2' \psi ' \mathcal{F}'-2 \mathcal{F}^{3/2} \left(f a_4''+a_4' f'\right)\Big)+a_2 \Big(-2 \mathcal{F}^{3/2} \left(f a_2''+a_2' f'\right)\nonumber\\
&&-\sqrt{D-2}r f a_4' \psi ' \mathcal{F}'+2 \sqrt{D-2} \mathcal{F} \left(r f a_4' \psi ''+\psi ' \left(r f a_4''+a_4' \left(r f'+f\right)\right)\right)\Big)\Bigg]\ .
\label{pentential U-}
\eea

\end{document}